\documentclass[twocolumn,secnumarabic,amssymb, nobibnotes, aps, prd]{revtex4-2}

\usepackage{amsmath} 
\usepackage{mathtools} 
\usepackage{caption}
\usepackage{subcaption}
\usepackage[T1]{fontenc}
\usepackage[latin9]{inputenc}
\usepackage{hyperref}

\begin{document}

\title{Calculating the many-potential vacuum polarization density of the Dirac equation in the finite-basis approximation}
\author{Maen Salman}
\email[Electronic address: ]{msalman@irsamc.ups-tlse.fr}
\author{Trond Saue}
\affiliation{Laboratoire de Chimie et Physique Quantique\\ UMR 5626 CNRS-Université Toulouse III-Paul Sabatier\\ 118 Route de Narbonne, F-31062 Toulouse, France}

\begin{abstract}
In this work, we propose an efficient and accurate computational method to evaluate the many-potential $\alpha\left(Z\alpha\right)^{n\ge3}$ vacuum polarization density of hydrogen-like atoms within the finite-basis approximation of the Dirac equation. To prove the performance of our computational method, we choose to work with the one-electron $_{\,\,\,92}^{238}\text{U}$ atom. In summary, we find that compliance with charge conjugation symmetry is a priori required to obtain physical results that are in line with our knowledge of the analytical problem.
We also note that the final numerical results are found to be in excellent agreement with previous formal analytical (and numerical) evaluations that are limited to a few simple nuclear distribution models. Our technique can be efficiently implemented and evaluated in codes that solve the radial Dirac equation in the finite basis set framework and allows the use of arbitrary (radial) nuclear charge distribution. The obtained numerical results of the non-perturbative vacuum polarization density automatically account for the extended nuclear size effect. This method is hence of special importance for atomic Dirac problems whose analytical Green's functions expressions are not at hand or have relatively complicated analytical forms. Furthermore, we propose a vacuum polarization density formula that forces compliance with charge conjugation symmetry and can be used in cases where the relativistic basis violates this symmetry, as is the case in most relativistic basis set programs. In addition, we have shown that vector components of the vacuum polarization four-current vanish in the case where the Dirac Hamiltonian is symmetric under time-reversal symmetry.
\end{abstract}
\maketitle{}

\tableofcontents{}

\section{Introduction}

In the seminal work of Wichmann and Kroll \cite{Wichmann_Kroll1956}, the bound-state (Furry picture QED) vacuum polarization effect, of order $\alpha\left(Z\alpha\right)^{n\geq1}$, was rigorously studied for the point nucleus problem. Their technique lay in expressing the vacuum polarization (VP) density in terms of the trace of the two-by-two radial Dirac Green's function. In addition, they constructed their Green's function from two radial solutions of the Dirac equation (in the presence of a Coulomb potential), each satisfying the right boundary condition at the origin, and at infinity, respectively, through what is known in the differential equation framework as the Wronskian method. This problem was also studied by Hylton \cite{hylton1984reduced}, Yerokhin and Maiorova \cite{Yerokhin-Maiorova2022-Green-function}, Grant \cite[section 3.6]{Grant}, Swainson and Drake \cite{Swainson_Drake1991}, and Hill \cite{Hill2006}. The importance of this construction lies in the fact that it avoids using the conventional definition of the Green's function, given as a sum (and integration) of outer products over the whole set of Dirac solutions with poles placed on corresponding eigenvalues (spectral representation). After a lengthy derivation, Wichmann and Kroll obtained an exact expression for the total Laplace-transformed VP density and showed that the regularized version of the first-order contribution -- that is linear in the external-potential strength $Z\alpha$ -- represents the VP density associated with the Uehling potential \cite{Uehling1935}, and furthermore isolated the third-order $\left(Z\alpha\right)^{3}$ VP density. Later, Blomqvist \cite{BLOMQVIST197295} evaluated the inverse Laplace transform of this third-order density and derived the associated real-space  $\alpha\left(Z\alpha\right)^{3}$ VP effective potential, which solves, together with the associated density, the electrostatic Poisson equation. This potential can be directly employed in practical calculations to account for the third-order VP effect. Due to their mathematical complexity, effective potentials associated with the higher-order $\alpha\left(Z\alpha\right)^{n\geq5}$ VP problems were never derived. 

Up to this point, the external potential was assumed to be the one generated by a point nucleus. The finite nuclear size effect on the $\alpha\left(Z\alpha\right)^{n\ge3}$ VP correction was first computed by Rinker and Wilets \cite{RinkerWilets1973PhysRevLett.31.1559}, in addition to Gyulassy \cite{Gyulassy_PhD_1974}. 

In the former work, the authors computed the total VP density generated by a finite nucleus charge density (Fermi distribution); their VP density expression included a finite sum over bound states and a numerical integration along continuum solutions (up to some large enough momentum). The obtained many-potential VP density results suffered from the following non-physical aspects: 1) It contains a finite gauge-noninvariant contribution that needs to be removed, known from the simplest photon-photon scattering process; this aspect is discussed in Refs. \cite{RinkerWilets1973PhysRevLett.31.1559},\cite[pages 40-42 and Appendix I]{Gyulassy_PhD_1974},  \cite{Rinker_Wilets1975PhysRevA.12.748}, \cite[section III.A.3]{Borie_Rinker1982RevModPhys.54.67}, and \cite{SoffMohr1988PhysRevA.38.5066}. 2) It does not integrate to zero. The authors then employed further treatments to refine their numerical results.

In the latter (more rigorous) work, Gyulassy \cite{Gyulassy_PhD_1974} followed Wichmann and Kroll and constructed the radial Green's function associated with a general extended nuclear distribution of a definite charge radius $r_{\text{n}}$ (this is the case of the sphere- and ball-nuclei) using the following reasoning. The solution that is regular at infinity (far away from the nucleus) is the one that solves the point nucleus problem. On the other hand, the solution that is regular at zero (inside the nucleus) solves the free particle equation, in the shell nucleus case. The full radial solution is then constructed from these two solutions, and by imposing wavefunction continuity at $r=r_{\text{n}}$. Gyulassy considered both sphere (shell) and ball nuclei, proceeded in computing the individual radial Green's function associated with $Z\alpha$ and $(Z\alpha)^3$ VP orders, calculated their corresponding individual VP densities, and provided an estimate of the extended nuclear distribution effect on the VP density generated by a point nucleus.

For further studies of the many-potential VP effect in the presence of a finite nucleus, the reader may consult Refs. \cite{Rinker_Wilets1975PhysRevA.12.748,BrownCahnMcLerran1975PhysRevD.12.609,Borie_Rinker1982RevModPhys.54.67,Neghabian1983_PhysRevA.27.2311,SoffMohr1988PhysRevA.38.5066,SchmidtSoffMohr1989PhysRevA.40.2176,LEE_Milstein199472,SapirsteinCheng2003PhysRevA.68.042111,soff1989influence,mohr1998qed}. We finally note that an alternative computation of this many-potential VP effect, in the presence of an arbitrary radial nuclear distribution, was provided by Persson \textit{et al.} \cite{PersonLindgrenSalomonsonSunnergren1993PhysRevA.48.2772} (see also Grant and Quiney \cite{GrantQuiney2022atoms10040108}, as well as Sunnergren \cite{sunnergren1998Thesis}). Their construction considered removing the linear $\alpha\left(Z\alpha\right)$ contribution (one-potential) from the total VP potential rather than its associated VP density, through a partial-wave-expansion (decomposition) technique. This technique allows to decompose the full problem into individual $\pm\kappa$ problems, yields a term-by-term divergence cancellation between contributions of the opposite sign of $\kappa$, forces the spurious finite gauge-noninvariant contribution to vanish, and yields accurate numerical results, as observed by Soff and Mohr \cite[page 5068]{SoffMohr1988PhysRevA.38.5066}.

In this work, on the other hand, we shall tackle the problem from a totally different angle, avoiding the numerical implementation of (relatively) complicated analytical expressions and their associated numerical integrations. Our primary motivation is the efficient computation of the VP density, specifically in the framework of the finite-basis approximation of the Dirac equation \cite[section 22.6]{Grant2006_chapter}, yet without loss of numerical precision. The importance of performing such calculations is that it is not based on a radial discretization of the spherical space centered at the nuclear position (suited for atomic calculations) \cite[section 22.6]{Grant2006_chapter}, but rather assume that the exact radial solution can be adequately described by a linear combination of a finite set of radial functions (basis set functions) that 1) satisfy physical boundary condition requirements and 2) (if it is possible/practical) follow the behavior of the exact wavefunction. This technique is widely used in both atomic and molecular calculations, where in the latter case a set of basis functions is centered on each of the nuclear positions. Furthermore, in our calculations, we shall consider Gaussian-type basis functions whose physical and mathematical significance is discussed in Sec. \ref{subsec:Gaussian-basis-functions}.

After constructing the matrix representation associated with the Dirac Hamiltonian, we numerically compute its eigensolutions and proceed to calculate the VP density, from the obtained solutions, using the conventional VP density definition, which takes the difference between positive- and negative-energy one-electron-state charge densities. We next subtract the linear term (in $Z\alpha$), containing the physical Uehling correction together with a non-physical divergence, through a simple procedure, proposed by Rinker and Wilets \cite{RinkerWilets1973PhysRevLett.31.1559,Rinker_Wilets1975PhysRevA.12.748}, and show that the obtained non-perturbative (many-potential) VP density results are in excellent agreement with the previous results of Mohr \textit{et al.} \cite[section 4.2]{mohr1998qed} that concerned the one-electron uranium atom, where the nucleus was represented by a sphere-like nuclear distribution (hollow sphere model).%

We note that the efficiency of our method lies in the fact that it avoids any kind of numerical integration (as done in the work of Gyulassy \cite{Gyulassy_PhD_1974}, Rinker and Wilets \cite{RinkerWilets1973PhysRevLett.31.1559}, or Persson \textit{et al.} \cite{PersonLindgrenSalomonsonSunnergren1993PhysRevA.48.2772}) in computing the VP density. It can also be shown that the spurious gauge-noninvariant contact term, discussed in Refs. \cite{RinkerWilets1973PhysRevLett.31.1559,Gyulassy_PhD_1974}, and \cite[section III.3]{Borie_Rinker1982RevModPhys.54.67}, automatically vanishes in our calculation. This is a direct consequence of the fact that in a finite basis framework, the Green's function that is constructed out of the finite basis solutions is no longer singular in the limit of coinciding spatial points, in addition to the fact that the obtained solutions form a complete orthogonal eigenbasis of finite size.

Furthermore, due to the kinetic balance condition, one obtains a total even number of eigenvalues that equally splits between positive and negative eigenvalues, as indicated by Stanton and Havriliak \cite{stanton1984kinetic}. Moreover, since our solutions are normalized, the use of kinetic balance implies that the total VP charge (spatial integral of the VP density) must always vanish. Finally, in the case where charge conjugation symmetry (${\cal C}$-symmetry) is realized in the finite basis set, all even orders of interaction with the external potential vanish, as recently indicated by Grant and Quiney \cite{GrantQuiney2022atoms10040108}. This reasoning goes back to Furry \cite{Furry1937theorem} who used ${\cal C}$-symmetry to prove that there should be no physical contributions coming from QED corrections that are represented by Feynman diagrams containing closed free-electron loops with an odd number of vertices. All expressions used and developed in this work are written in SI units in order to facilitate their conversion to the favorite choice of units adopted by the reader.

\section{Theory}

\label{sec:Theory}

The existence of an external non-quantized current source $J^{\text{ext.}}=\left(c\rho^{\text{ext.}},\boldsymbol{J}^{\text{ext.}}\right)$ in vacuum, where $\rho^{\text{ext.}}$ is the volume charge density and $\boldsymbol{J}^{\text{ext.}}$ is the volume current density, polarizes the electron-positron pairs that are simultaneously created from the vacuum, and annihilated into it. The collective emergence of these pairs forms what is known as the VP density cloud, which surrounds the inducing source, and screens its interaction with other particles. In the atomic problem, where the nucleus is typically assumed to be spherically symmetric, a spherical VP cloud forms inside and closely around the nucleus, and screens its Coulombic interaction with orbiting bound electrons.

In Lorenz gauge, the four-potential generated by the external source satisfies the following non-homogeneous Maxwell's equations
\begin{equation}
\begin{aligned}\square A_{\mu}^{\text{ext.}}\left(x\right) & =\mu_{0}J_{\mu}^{\text{ext.}}\left(x\right),\\
\text{with }\,\,\, \square & :=\partial^{\mu}\partial_{\mu}=\frac{1}{c^{2}}\frac{\partial^{2}}{\partial t^{2}}-\boldsymbol{\nabla}^{2}.
\end{aligned}
\end{equation}

Here, the four-position and four-gradient are given by $x^\mu=\left(ct,\boldsymbol{x}\right)$ and $\partial_\mu=\partial/\partial x^{\mu}$, respectively. The external four-potential $A^{\text{ext.}}=\left(\phi^{\text{ext.}}/c,\boldsymbol{A}^{\text{ext.}}\right)$ contains the scalar potential $\phi^{\text{ext.}}$ in addition to the (magnetic) vector-potential $\boldsymbol{A}^{\text{ext.}}$. In the (static) atomic problem, the scalar potential energy can be written as
\begin{equation}
-e\phi^{\text{ext}.}\left(x\right)=-\left(Z\alpha\right)\hbar c\int d^{3}y\frac{\rho^{\text{n}}\left(\boldsymbol{y}\right)}{\left|\boldsymbol{x}-\boldsymbol{y}\right|},\label{eq:external-potential}
\end{equation}
where $-e$ is the electron charge, $\alpha=e^{2}/4\pi\epsilon_{0}\hbar c$ is the fine structure constant, and $\rho^{\text{n}}$ is some arbitrary normalized nuclear distribution. The VP four-current generated by the external source can be written as \cite[Eq.(2.11)]{SchwingerPhysRev.82.664_1951}
\begin{equation}
J_{\mu}^{\text{VP}}\left(x\right)=i\hbar ec\text{Tr}\left[\gamma_{\mu}S_{A}^{F}\left(x,y\right)\right]_{y\rightarrow x},\label{eq:J_mu}
\end{equation}
where $S_{A}^{F}\left(x,y\right)$ is the Feynman propagator of the Dirac problem \cite[Eq.(17)]{Feynamn_1949_Positron} that satisfies
\begin{equation}
\big[\gamma^{\mu}\big(i\hbar\partial_\mu+eA_{\mu}^{\text{ext.}}\left(x\right)\big)-mc\big]S_{A}^{F}\left(x,y\right)=\delta^{4}\left(x-y\right),
\end{equation}
and which can be written as a vacuum expectation value of the time-ordered product of two electron field operators. This corresponds to the Feynman choice of energy contour integration that enters the inverse Fourier transform expression of $S_{A}^{F}$, and defines the Feynman propagator. Discussions about the Feynman propagator in the presence of an external potential can be found in Refs. \cite[sections 2.5 and 3.1.4]{Itzykson_Zuber_QFT_1980}, \cite[section 15g]{schweber2011introduction} and \cite[chapter 2]{greiner_reinhard2009QED}. 
Note also that Schwinger points out that the space-time limit in Eq.(\ref{eq:J_mu}) should be taken symmetrically with respect to past and future (see, instance, discussion in Ref. \cite[Eq.(9.111) and section 14.1]{Greiner1985QEDstrong}).

We choose the $\gamma^{\mu}=\left(\gamma^{0},\boldsymbol{\gamma}\right)$ matrices to be those associated with the Dirac representation. The $Z\alpha$ factor in Eq.(\ref{eq:external-potential}) describes the nuclear potential strength and shall be used as an expansion parameter for VP quantities throughout this work. In the case where the source current $J^{\text{ext.}}$ is time-independent, the general VP current expression reduces to the following time-independent expression
\begin{equation}
\begin{aligned} & J_{\mu}^{\text{VP}}\left(x\right)=J_{\mu}^{\text{VP}}\left(\boldsymbol{x}\right)\\
 & =\frac{ec}{2}\big[\sum_{E_{n}>0}\bar{\psi}_{n}\left(\boldsymbol{x}\right)\gamma_{\mu}\psi_{n}\left(\boldsymbol{x}\right)\\
 & -\sum_{E_{n}<0}\bar{\psi}_{n}\left(\boldsymbol{x}\right)\gamma_{\mu}\psi_{n}\left(\boldsymbol{x}\right)\big],
\end{aligned}\label{eq:VP-current}
\end{equation}
where $\bar{\psi}_{n}=\psi_{n}^{\dagger}\gamma^{0}$ is the Dirac adjoint \cite{peierls1934vacuum}, and where $\psi_{n}\left(\boldsymbol{x}\right)$ and $E_{n}$ form a solution of the time-independent Dirac equation
\begin{equation}
\begin{aligned}H & \psi_{n}\left(\boldsymbol{x}\right)=E_{n}\psi_{n}\left(\boldsymbol{x}\right)\\
H & =c\boldsymbol{\alpha}\cdot[-i\hbar\boldsymbol{\nabla}+e\boldsymbol{A}^{\text{ext.}}\left(\boldsymbol{x}\right)]+\beta mc^{2}-e\phi^{\text{ext.}}\left(\boldsymbol{x}\right),
\end{aligned}\label{eq:Dirac-indep}
\end{equation} 
in the presence of the time-independent source; here, $\boldsymbol{\alpha}=\gamma^{0}\boldsymbol{\gamma}$ and $\beta=\gamma^{0}$ are the conventional Dirac matrices. We note that Eq.(\ref{eq:VP-current}) is formal; discrete sums over positive- and negative-energy continuum solutions are to be replaced by corresponding integrals over energy-continua.

In the special case where the nuclear potential is assumed to be spherically symmetric, it was shown by Indelicato \textit{et al.} \cite{Indelicato_Mohr_SapirsteinCoordinateSpaceVP2014} that the vector components of the VP four-current vanishes. However, a more general statement about the vanishing of the  VP current density $\boldsymbol{J}^{\text{VP}}$ can be given by consideration of time-reversal symmetry \cite[section 2.8.2]{salman2022quantum}. The time-reversal operator is given by
\begin{equation}
{\cal T}=U_{T}{\cal K}_{0},
\end{equation}
where ${\cal K}_{0}$ is the complex conjugation operator, and $U_{T}=\gamma^{1}\gamma^{3}$ is the unitary matrix operator associated with the time-reversal operation. In the absence af an external vector potential, $\boldsymbol{A}^{\text{ext.}}=0$, each wavefunction $\psi_{n}\left(\boldsymbol{x}\right)$, associated with energy $E_{n}$, has a (Kramers) partner ${\cal T}\psi_{n}\left(\boldsymbol{x}\right)$ with the same energy $E_{n}$ (see, for instance, Ref. \cite[section 11.4]{schwabl2007quantum}). The contribution of a solution and its time-reversed partner to components of the VP four-current may be expressed as
\begin{equation}
\psi_{n}^{\dagger}\left(\boldsymbol{x}\right)\gamma^{0}\gamma^{\mu}\psi_{n}\left(\boldsymbol{x}\right)+\psi_{n}^{\dagger}\left(\boldsymbol{x}\right)\left[U_{T}^{\dagger}\gamma^{0}\gamma^{\mu}U_{T}\right]^{t}\psi_{n}\left(\boldsymbol{x}\right).
\end{equation}

From inspection of the sandwiched matrix of the second term, one finds that
\begin{equation}
\left[U_{T}^{\dagger}\gamma^{0}\gamma^{\mu}U_{T}\right]^{t}=\begin{cases}
1 & \text{if }\mu=0\\
-\gamma^{0}\gamma^{\mu} & \text{otherwise}
\end{cases},
\end{equation}
showing that the vector component ($\mu=1,2,3$) of Eq.(\ref{eq:VP-current}) vanishes in the time-symmetric Dirac problem. On the other hand, the time-component of the VP four-current, the VP charge density
\begin{equation}
\begin{aligned}\rho^{\text{VP}}\left(\boldsymbol{x}\right) & =\frac{e}{2}\big[\sum_{E_{n}>0}\psi_{n}^{\dagger}\left(\boldsymbol{x}\right)\psi_{n}\left(\boldsymbol{x}\right)\\
 & -\sum_{E_{n}<0}\psi_{n}^{\dagger}\left(\boldsymbol{x}\right)\psi_{n}\left(\boldsymbol{x}\right)\big].
\end{aligned},\label{eq:vac-pol-den}
\end{equation}
is generally non-zero. Although this equation, to our knowledge, appears for the first time in the work of Wichmann and Kroll \cite[Eq.(2)]{Wichmann_Kroll1956}, one finds its roots in the work of Dirac \cite{dirac_1934} that concerned relativistic density matrices and their associated divergences, in addition to the works of Schwinger of Refs. \cite[Eq.(1.14)]{SchwingerPhysRev.74.1439_1948}, \cite[Eq.(1.69)]{SchwingerPhysRev.75.651_1949} and \cite[Eqs.(2.3 and 2.10)]{SchwingerPhysRev.82.664_1951} in which the VP current of Eq.(\ref{eq:VP-current}) is employed. 

Another symmetry that shall be of particular importance in this work is ${\cal C}$-symmetry; the symmetry that connects the electron (particle) quantum state to the one associated with its anti-particle partner (positron). For instance, when there are no external sources, that is, when both $\boldsymbol{A}^{\text{ext.}}$ and $\phi^{\text{ext.}}$ are zero, corresponding to the free-particle case, one can show, using ${\cal C}$-symmetry, that the VP charge density $\rho^{\text{VP}}$ vanishes as well. The ${\cal C}$ operator can be written as \cite[section 11.3]{schwabl2007quantum}
\begin{equation}
{\cal C}=U_{C}{\cal K}_{0},\quad\text{with}\quad U_{C}=\gamma^{2}.
\end{equation}

Using this operator, one can relate opposite energy-sign solutions, of Eq.(\ref{eq:Dirac-indep}), through $\psi_n^\pm (\boldsymbol{x}) = {\cal C}\psi_n^\mp(\boldsymbol{x})$, where $+$ and $-$ superscripts are added to distinguish between positive- and negative-energy free-solutions, respectively. As a consequence, we can write the density associated with a free positive-energy solution as
\begin{equation}
\begin{aligned}  \psi_{n}^{+\dagger}\left(\boldsymbol{x}\right)\psi_{n}^{+}\left(\boldsymbol{x}\right) & =\psi_{n}^{-\dagger}\left(\boldsymbol{x}\right)U_{C}^{\dagger}U_{C}\psi_{n}^{-}\left(\boldsymbol{x}\right)\\
 & =\psi_{n}^{-\dagger}\left(\boldsymbol{x}\right)\psi_{n}^{-}\left(\boldsymbol{x}\right).
\end{aligned}
\end{equation}

This result shows that every positive-energy density contribution is balanced by a negative-energy density contribution, yielding a vanishing total vacuum polarization density in Eq.(\ref{eq:vac-pol-den}).

Before closing this section, we stress that the vacuum polarization current of Eq.(\ref{eq:J_mu}) is strictly divergent. This is due to the fact that the Feynman propagator (and the Dirac Green's function) diverges in the limit of coinciding space-time points ($y\rightarrow x$) and implies a divergent vacuum polarization density. This real-space problem was considered by Indelicato \textit{et al.} \cite{Indelicato_Mohr_SapirsteinCoordinateSpaceVP2014} using the Pauli-Villars regularization scheme, where auxiliary-mass propagators are introduced to regularize divergent quantities, in an approach that is similar to the conventional Fourier space treatment. 

In this work, we shall consider the finite-basis approximation of the radial Dirac equation, where divergences can only be manifested by finite spurious (non-physical) contributions. These contributions must be eliminated in order to obtain valid physical results. 

\subsection{Radial Dirac problem}

We shall now focus on the case where the external scalar potential is spherically symmetric. In this case, the Dirac spinor can then be written as \cite[section 2.6]{johnson2007atomic}
\begin{equation}
\psi_{n,\kappa,m_{j}}\left(\boldsymbol{x}\right)=\frac{1}{r}\begin{bmatrix}P_{n,\kappa}\left(r\right)\Omega_{\kappa,m_{j}}\left(\hat{\boldsymbol{x}}\right)\\
iQ_{n,\kappa}\left(r\right)\Omega_{-\kappa,m_{j}}\left(\hat{\boldsymbol{x}}\right)
\end{bmatrix},\label{eq:spherical-Dirac-sol}
\end{equation}
where $r=|\boldsymbol{x}|$ is the radial distance, $n$ is the principal quantum number, $\kappa$ is the relativistic angular quantum number \cite[section 1.5]{johnson2007atomic}, and $m_{j}$ is the secondary total angular momentum quantum number. $\Omega_{\kappa,m_j}$ is the two-component spherical spinor. $P_{n,\kappa}$ and $Q_{n,\kappa}$ are large and small component radial functions associated with the $E_{n,\kappa}$ energy level. These three quantities form a solution of the radial Dirac equation
\begin{equation}
\left(h_{\kappa}-E_{n,\kappa}\right)\varphi_{n,\kappa}=0,\label{eq:radial-Dirac}
\end{equation}
where the radial Dirac Hamiltonian is given by
\begin{equation}
h_{\kappa}=\begin{bmatrix}mc^{2}-e\phi^{\text{ext}.}\left(r\right) & -c\hbar\left[\frac{d}{dr}-\frac{\kappa}{r}\right]\\
c\hbar\left[\frac{d}{dr}+\frac{\kappa}{r}\right] & -mc^{2}-e\phi^{\text{ext}.}\left(r\right)
\end{bmatrix},\label{eq:radial-Dirac-Hamiltonian}
\end{equation}
with the corresponding two-component radial solution
\begin{equation}
\varphi_{n,\kappa}=\begin{bmatrix}P_{n,\kappa}\\
Q_{n,\kappa}
\end{bmatrix}\label{eq:radial-Dirac-spinor}.
\end{equation}

If we now plug the relativistic atomic orbital of Eq.(\ref{eq:spherical-Dirac-sol}) into the VP density expression of Eq.(\ref{eq:vac-pol-den}), and sum the product of spherical spinors over $m_{j}$ (using \cite[Eq.(3.12)]{Szmytkowski_2005}), we obtain the following VP density expression \cite[Eq.(8)]{Wichmann_Kroll1956}
\begin{align}
\rho^{\text{VP}}\left(\boldsymbol{x}\right) & =\sum_{\kappa=\pm1,\pm2\ldots}\rho_{\kappa}^{\text{VP}}\left(\boldsymbol{x}\right)\\
\rho_{\kappa}^{\text{VP}}\left(\boldsymbol{x}\right) & =\frac{e\left|\kappa\right|}{4\pi}\frac{1}{r^{2}}\sum_{n}\text{sgn}\left(E_{n,\kappa}\right)\rho_{n,\kappa}(r),\label{eq:VP-kappa}
\end{align}
where $\rho_{n,\kappa}=\varphi_{n,\kappa}^{\dagger}\varphi_{n,\kappa}$ is the radial probability density associated with the $E_{n,\kappa}$ energy-level.%

In the case of an atom of charge $Z$, one can verify that the radial electronic $h_{Z,\kappa}$ and positronic $h_{-Z,-\kappa}$ Hamiltonians are related through 
\begin{equation}
\sigma_{1}h_{Z,\kappa}\sigma_{1}=-h_{-Z,-\kappa},
\end{equation}
where $\sigma_1=\begin{bsmallmatrix}0&1\\1&0\end{bsmallmatrix}$ is the first Pauli matrix. We note that having a negative sign of $Z$ means that the Dirac electron interacts with a negative nuclear charge, and this is equivalent to having a Dirac positron interacting with a positive nuclear charge. If we next assume that $E_{Z,n,\kappa}$ and $\varphi_{Z,n,\kappa}$ are eigensolutions of the electronic problem, i.e., they solve the time-independent radial equation
\begin{equation}
h_{Z,\kappa}\varphi_{Z,n,\kappa}=E_{Z,n,\kappa}\varphi_{Z,n,\kappa},
\end{equation}
then, one can show that the positronic eigensolutions $E_{-Z,n,\kappa}$
and $\varphi_{-Z,n,\kappa}$, solving the corresponding positronic Hamiltonian $h_{-Z,\kappa}$
equation, are related to the electronic solutions through the following relations
\begin{align}
E_{-Z,n,\kappa} & =-E_{Z,n,-\kappa}\\
\varphi_{-Z,n,\kappa} & =\sigma_{1}\varphi_{Z,n,-\kappa}.\label{eq:C-symmetry-eigen}
\end{align}

It should become clear now that in the absence of an external potential
($Z=0$), the last relations reduce to \cite{Csymm_salman_saue}
\begin{align}
E_{0,n,\kappa} & =-E_{0,n,-\kappa}\\
\varphi_{0,n,\kappa} & =\sigma_{1}\varphi_{0,n,-\kappa},\label{eq:C-symmetry-eigenZ0}
\end{align}
showing that ${\cal C}$-symmetry connects free-particle eigensolutions
of opposite sign of energy and $\kappa$ quantum number. Using these
relations, our VP density associated with the $-\kappa$ problem can
be written as \cite[section  4.4]{salman2022quantum}
\begin{equation}
\begin{aligned}\rho_{-\kappa}^{\text{VP}}\left(\boldsymbol{x}\right) & =-\frac{e\left|\kappa\right|}{4\pi}\frac{1}{r^{2}}\sum_{n}\text{sgn}\left(E_{0,n,\kappa}\right)\rho_{0,n,\kappa}\left(r\right)\\
 & =-\rho_{+\kappa}^{\text{VP}}\left(\boldsymbol{x}\right),\label{eq:VP-kappa-kappa}
\end{aligned}\end{equation}
where $\rho_{0,n,\kappa}=\varphi_{0,n,\kappa}^{\dagger}\varphi_{0,n,\kappa}$.
This relation shows that the total VP density of Eq.(\ref{eq:vac-pol-den}) vanishes due to a total cancellation between opposite $\kappa$ sign radial VP densities.
In the atomic problem (where $Z\neq0$), a partial cancellation between these contributions is expected, as indicated in Refs. \cite[section 5.4]{sunnergren1998Thesis}, \cite{PersonLindgrenSalomonsonSunnergren1993PhysRevA.48.2772}, and \cite[section 4.5.3]{salman2022quantum}, and to be shown in the numerical Sec. \ref{sec:Numerical-computations}. For this reason, one should always compute the radial VP density in pairs of opposite $\kappa$-sign contributions
\begin{equation}
\rho_{\left|\kappa\right|}^{\text{VP}}\left(\boldsymbol{x}\right)=\rho_{+\kappa}^{\text{VP}}\left(\boldsymbol{x}\right)+\rho_{-\kappa}^{\text{VP}}\left(\boldsymbol{x}\right).
\end{equation}

Furthermore, we shall show that in practical radial calculations where we set $Z=0$ (the free-particle problem), if the ${\cal C}$-symmetry is not realized, Eq.(\ref{eq:VP-kappa-kappa}) will not hold and as a consequence, the free VP density shall not vanish. This is a worrying non-physical result.

\subsection{Dirac equation in finite basis}

\label{subsec:Dirac-equation-in}

The basic idea behind the (relativistic) finite-basis method is to approximate large and small component radial functions $P_{n,\kappa}$ and $Q_{n,\kappa}$ by a finite set of basis functions, that is (by construction) respecting the right radial boundary conditions of the exact radial functions, at both short and large distances. This machinery allows the transformation of the differential eigenvalue problem into an integral eigenvalue problem, which can be readily solved on a computer. We start by introducing a four-component basis set (two radial components) in which the radial Dirac spinor of Eq.(\ref{eq:radial-Dirac-spinor}) is expanded as
\begin{equation}
\varphi_{\alpha,\kappa}=\sum_{i=1}^{n_{\kappa}}c_{\alpha,\kappa,i}\begin{bmatrix}\pi_{\kappa,i}^{+}\\
\pi_{\kappa,i}^{-}
\end{bmatrix},\label{eq:4c-bases}
\end{equation}
where $\pi_{\kappa,i}^{\pm}$ are some large $(+)$ and small $(-)$ radial functions, and $n_\kappa$ represents the basis set size of the $\kappa$-problem. Alternatively, one can expand the Dirac spinor by two independent large and small sets of basis functions
\begin{equation}
\varphi_{\alpha,\kappa}=\sum_{i=1}^{n_{\kappa}^{+}}c_{\alpha,\kappa,i}^{+}\begin{bmatrix}\pi_{\kappa,i}^{+}\\
0
\end{bmatrix}+\sum_{i=1}^{n_{\kappa}^{-}}c_{\alpha,\kappa,i}^{-}\begin{bmatrix}0\\
\pi_{\kappa,i}^{-}
\end{bmatrix}.\label{eq:22c-bases}
\end{equation}

Early representations of the Dirac equation in the finite basis set framework suffered from the appearance of spurious eigenvalues, and the occurrence of variational collapse \cite{kutzelnigg1984basis,tupitsyn_Shabaev_2008spurious,Goldman1985}. For a detailed mathematical study on the occurence of spurious solutions in different relativistic basis sets the reader can consult the more recent works of Lewin and S\'er\'e \cite{Lewin_Sere_2009_spectral-pollution,lewin2014spurious}.

It was later found that the reason behind this instability was that the same (or arbitrary) set of basis functions was given for $\pi_{\kappa,i}^{+}$ and $\pi_{\kappa,i}^{-}$, while from the Dirac equation of Eq.(\ref{eq:radial-Dirac}), we see that large and small components are coupled. To overcome 1) the observed unphysical results, and 2) the fact that the exact coupling between the radial components is energy-dependent (unknown before computation), the kinetically balanced (KB) basis was introduced \cite{Schwarz_and_Wallmeir_1982,Grant_1982,stanton1984kinetic,Dyall_1984_matrix_operators,kutzelnigg2007completeness}. Following the KB prescription, which is valid for positive-energy solutions, one uses the basis expansion of Eq.(\ref{eq:22c-bases}), introduces a set of large component radial functions $\pi_{\kappa,i}^{+}$ and generates the small component radial functions through $\pi_{\kappa,i}^{-}=\frac{\hbar}{2mc}\left[\frac{d}{dr}+\frac{\kappa}{r}\right]\pi_{\kappa,i}^{+}$. This coupling between components is obtained from the exact coupling after assuming that 1) the energy can be approximated by $E\approx mc^{2}$, which holds (to some extent) for bound-states, and 2) the external potential can be neglected in front of this energy $\left|e\phi^{\text{ext}}\right|\ll mc^{2}$, which is obviously not valid for point nuclei. %
Similarly, one can consider the negative-energy version of this prescription, assume that the energy can be approximated as $E\approx -mc^2$, and that the scalar potential can be neglected in front of this energy. This reasoning leads to the inverse kinetic balance (IKB) \cite{sun2011comparison} basis construction, where one uses the basis expansion of Eq.(\ref{eq:22c-bases}), introduces small component radial functions $\pi^{-}_{\kappa,i}$, and generates the large component basis function using $\pi_{\kappa,i}^{+}=\frac{\hbar}{2mc}\left[\frac{d}{dr}-\frac{\kappa}{r}\right]\pi_{\kappa,i}^{-}$.

Finally, a more symmetric treatment between positive- and negative-energy
solutions is provided by the dual kinetic balance (DKB) prescription,
proposed by Shabaev \textit{et al.} in Ref. \cite{Shabaev_2004_DKB},
in which one writes
\begin{equation}
\begin{aligned}\varphi_{\alpha,\kappa}^{\text{DKB}} & =\sum_{i=1}^{n_{\kappa}^{+}}c_{\alpha,\kappa,i}^{+}\begin{bmatrix}\pi_{\kappa,i}^{+}\\
\frac{\hbar}{2mc}\left[\frac{d}{dr}+\frac{\kappa}{r}\right]\pi_{\kappa,i}^{+}
\end{bmatrix}\\
 & +\sum_{i=1}^{n_{\kappa}^{-}}c_{\alpha,\kappa,i}^{-}\begin{bmatrix}\frac{\hbar}{2mc}\left[\frac{d}{dr}-\frac{\kappa}{r}\right]\pi_{\kappa,i}^{-}\\
\pi_{\kappa,i}^{-}
\end{bmatrix}.
\end{aligned}\label{eq:DKB}
\end{equation}

It should be noted, however, that within both KB and IKB prescriptions, following the basis construction of Eq.(\ref{eq:22c-bases}), large and small basis functions are decoupled, and this provides good computational flexibility. On the other hand, the DKB construction (following Eq.(\ref{eq:4c-bases})) combines both prescriptions while keeping the radial couplings between large and small component functions fixed. Similar to KB and IKB cases, the DKB construction is, in principle, not valid for the point nucleus problem. 

A recent interesting DKB-like scheme was proposed by Grant and Quiney \cite{GrantQuiney2022atoms10040108}, where the radial Dirac spinor is written as
\begin{equation}
\varphi_{\alpha,\kappa}^{\text{CKG}}=\sum_{i=1}^{n_{\kappa}^{+}}c_{\alpha,\kappa,i}^{+}\Phi_{\kappa,i}^{+}+\sum_{i=1}^{n_{\kappa}^{-}}c_{\alpha,\kappa,i}^{-}\Phi_{\kappa,i}^{-}.\label{eq:Quiney-DKB}
\end{equation}
Positive- and negative-energy basis elements are given by
\begin{equation}
\begin{aligned}\Phi_{\kappa,i}^{+} & =N_{\kappa,i}^{+}\begin{bmatrix}\pi_{\kappa,i}^{+}\\
\frac{\hbar}{mc+E_{\kappa,i}^{+}/c}\left[\frac{d}{dr}+\frac{\kappa}{r}\right]\pi_{\kappa,i}^{+}
\end{bmatrix}\\
\Phi_{\kappa,i}^{-} & =N_{\kappa,i}^{-}\begin{bmatrix}\frac{\hbar}{mc-E_{\kappa,i}^{-}/c}\left[\frac{d}{dr}-\frac{\kappa}{r}\right]\pi_{\kappa,i}^{-}\\
\pi_{\kappa,i}^{-}
\end{bmatrix},
\end{aligned}
\end{equation}
where $N_{\kappa,i}^{\pm}$ are the corresponding appropriate normalization constants. The radial couplings of the last basis elements are obtained from the exact couplings after assuming $\left|e\phi^{\text{ext}}\right|\ll E_{\kappa,i}^{\pm}$. As noted by Grant and Quiney, this choice of basis functions follows the exact coupling between large and small component function in the free-particle problem, where $\phi^{\text{ext}.}\left(r\right)=0$.
In addition, the energy parameters $E_{\kappa,i}^{\pm}$, are chosen
to be the unique positive and negative solutions to the following
equations
\begin{equation}
E_{\kappa,i}^{\pm}=\int_{0}^{\infty}dr\Phi_{\kappa,i}^{\pm\dagger}\left(r,E_{\kappa,i}^{\pm}\right)h_{\kappa}^{\text{Free}}\Phi_{\kappa,i}^{\pm}\left(r,E_{\kappa,i}^{\pm}\right),\label{eq:Epm}
\end{equation}
respectively, where $h_{\kappa}^{\text{Free}}$ is the free-particle version of the radial Dirac Hamiltonian $h_{\kappa}$ of Eq.(\ref{eq:radial-Dirac-Hamiltonian}). In the case where Gaussian basis functions (Eqs.(\ref{eq:pi_+} and \ref{eq:pi_-}) of Sec. \ref{subsec:Gaussian-basis-functions}) are employed, a straightforward calculation yields the following free-particle energy-momentum relation
\begin{align}
E_{\kappa,i}^{\pm} & =\pm c\sqrt{\langle p_{\kappa,i}^2 \rangle^{\pm}+m^{2}c^{2}}.\label{eq:Epm-energy-momentum}
\end{align}
The effective squared-momentum 
\begin{equation}
\langle p_{\kappa,i}^2 \rangle^{\pm}=\hbar^{2}\zeta^\pm_{\kappa,i}\left(|2\kappa\pm1|+2\right),
\end{equation} 
written in terms of the Gaussian exponent $\zeta^\pm_{\kappa,i}$ and the quantum number $\kappa$, is also directly obtained as an expectation value of the squared-momentum operator with respect to Gaussian basis functions $\pi^\pm_{\kappa, i}$, given in Eqs.(\ref{eq:pi_+} and \ref{eq:pi_-}). Eq.(\ref{eq:Epm-energy-momentum}) shows that each basis function $\Phi^{\pm}_{\kappa, i}$ is associated with a distinct energy parameter $E_{\kappa, i}^{\pm}$ which controls the coupling strength between radial components. 
We finally note that if the Gaussian basis is replaced by a Slater one, the energy parameter is found to be $\kappa$-independent
\begin{align}
E_{\kappa,i}^{\pm} & =\pm c\sqrt{\hbar^{2}(\zeta_{\kappa,i}^{\pm})^2+m^{2}c^{2}}.
\end{align}

For both bases, it is clearly seen that for small exponents $\zeta_{\kappa,i}^{\pm}\rightarrow0$, $E_{\kappa,i}^{\pm}\rightarrow\pm mc^{2}$, and the new construction of Eq.(\ref{eq:Quiney-DKB}) then coincides with the original DKB scheme of Shabaev \textit{et al.} \cite{Shabaev_2004_DKB}, given in Eq.(\ref{eq:DKB}). 


The proposal of Quiney and Grant is clearly interesting, for instance, showing excellent energy convergence for the atomic point nucleus problem, but shall not be further discussed or tested in the current work.

\subsection{${\cal C}$-symmetry in the finite basis}

\label{subsec:-symmetry-in-the-finite-basis}

In previous work, we considered the relativistic basis set compliance with ${\cal C}$-symmetry \cite{Csymm_salman_saue}. We have shown that the DKB construction can be made ${\cal C}$-symmetric if one forces the large and small basis functions to follow \cite[section 2.11.6]{salman2022quantum}
\begin{equation}
\pi_{\pm\kappa,i}^{+}=\pi_{\mp\kappa,i}^{-},
\end{equation}
and have concluded that the use of a Gaussian $j$-based basis sets assures such compliance; see Grant and Quiney \cite[Eq.(40)]{GrantQuiney2022atoms10040108}. This basis construction was discussed by Dyall \cite{DyallandFaegri1996}, where the same list of exponents is given for basis functions of same $j$ (total angular momentum) quantum number: basis functions of opposite signs of $\kappa$. We furthermore note that the ${\cal C}$-symmetry realization is achieved with a more general condition
\begin{equation}
\zeta_{\kappa,i}^{\pm}=\zeta_{-\kappa,i}^{\mp},\label{eq:zetas-C-DKB}
\end{equation}
which gives more flexibility for optimizing these exponents, since (in general) different sets of exponents can be given for large and small component Gaussians. Gaussian basis sets are discussed in section \ref{subsec:Gaussian-basis-functions}. This same analysis holds for the DKB construction of Grant and Quiney, discussed in the previous section.

In addition, we have considered the ${\cal C}$-symmetry realization in the KB and IKB problems. Here, we find that if the free-particle solutions (given in Ref. \cite[Eqs.(10,11)]{Csymm_salman_saue}, for instance) are used as basis set functions
\begin{align}
\pi_{\kappa,i}^{+} & =rj_{\left|\kappa+\frac{1}{2}\right|-\frac{1}{2}}\left(k_{\left|\kappa\right|,i}r\right)\\
\pi_{\kappa,i}^{-} & =rj_{\left|\kappa-\frac{1}{2}\right|-\frac{1}{2}}\left(k_{\left|\kappa\right|,i}r\right),
\end{align}
for the KB and IKB constructions, respectively, then the ${\cal C}$-symmetry is automatically realized.
We note that for the $\pm\kappa$ problems, one must introduce the same set of scaling factors $k_{\left|\kappa\right|,i}$ with $i=1,\ldots,n_{\kappa}$. To see how this realization is achieved, we proceed as follows. Using the spherical Bessel functions relations of Ref. \cite[Eqs.(10.1.21,22)]{Abramowitz+stegun}, one can directly write the small component function of the KB prescription as 
\begin{align*}
\left[\frac{d}{dr}+\frac{\kappa}{r}\right]\pi_{\kappa,i}^{+} & =+\text{sgn}\left(\kappa\right)k_{\left|\kappa\right|,i}\pi_{-\kappa,i}^{+}.
\end{align*}

This relation shows that the small component function of some $+\kappa$ problem (left-hand side) is a large component function of the $-\kappa$ problem (right-hand side), proving that this choice of basis is symmetric under ${\cal C}$-symmetry; cf. Eq.(\ref{eq:C-symmetry-eigenZ0}). Similarly, for the IKB problem, we find 
\begin{equation}
\left[\frac{d}{dr}-\frac{\kappa}{r}\right]\pi_{\kappa,i}^{-}=-\text{sgn}\left(\kappa\right)k_{\left|\kappa\right|,i}\pi_{-\kappa,i}^{-}.
\end{equation}

The main impractical feature of these basis sets is that there exists no closed expressions for the radial integrals, associated with the matrix representation of the radial Dirac equation.

\subsection{VP in the finite basis}

\label{subsec:VP-in-the-finite-basis}

We recall that in the radial problem, the VP density is written as (cf. Eq.(\ref{eq:VP-kappa})) 
\begin{align}
\rho_{\kappa}^{\text{VP}}\left(\boldsymbol{x}\right)=\frac{e\left|\kappa\right|}{4\pi}\frac{1}{r^{2}}\sum_{\alpha=1}^{n_{\kappa}}\text{sgn}\left(E_{\alpha,\kappa}\right)\rho_{\alpha,\kappa}\left(r\right),\label{eq:VP-in-basis}
\end{align}
where $\rho_{\alpha,\kappa}=\varphi_{\alpha,\kappa}^{\dagger}\varphi_{\alpha,\kappa}$
is now the radial probability density associated with the numerical solution of index $\alpha$ for a given $\kappa$ problem, and $n_{\kappa}$ represents the total number
of solutions, i.e., the basis set size. This density can be expanded
in powers of the nuclear charge $Z$ (the $Z\alpha$-expansion) as
\begin{equation}
\begin{aligned}\rho_{\kappa}^{\text{VP}}\left(\boldsymbol{x};Z\right) & =\sum_{n=0}^{\infty}\rho_{\kappa}^{\text{VP},n}\left(\boldsymbol{x};Z\right)\\
\rho_{\kappa}^{\text{VP},n}\left(\boldsymbol{x};Z\right) & =\frac{\partial^{n}}{\partial Z^{n}}\rho_{\kappa}^{\text{VP}}\left(\boldsymbol{x};Z\right)\Bigm{|}_{Z=0}\frac{Z^{n}}{n!}.
\end{aligned}\label{eq:VP-expansion}
\end{equation}

We know from Furry's theorem \cite{Furry1937theorem}, which is based on a ${\cal C}$-symmetry argument, that any free-electron loop with an odd number of vertices yields no physical contribution. This means that if the used basis set realizes ${\cal C}$-symmetry, then all even-order VP densities must vanish under ${\cal C}$-symmetry, as indicated by Wichmann and Kroll \cite[page 849]{Wichmann_Kroll1956}, and later by Gyulassy \cite[Eq.(2.19)]{Gyulassy_PhD_1974}. Furthermore, we know that the VP contribution that is linear in $Z$, which contains the (physical) Uehling contribution, is of an overall quadratic divergence (momentum space integration); the full Uehling contribution is obtained after summing over all possible values of $\kappa$. This degree of divergence is reduced to a logarithmic one once gauge-invariance (current conservation) is imposed on the polarization tensor; see for instance, \cite[section 15e]{schweber2011introduction}. In order to remove  this source of divergence, Rinker and Wilets \cite{Rinker_Wilets1975PhysRevA.12.748}, suggested eliminating the linear part of the VP density through the simple subtraction
\begin{align}
\rho_{\kappa}^{\text{VP},n\geq3}\left(\boldsymbol{x};Z\right) & =\rho_{\kappa}^{\text{VP}}\left(\boldsymbol{x};Z\right)-\rho_{\kappa}^{\text{VP},1}\left(\boldsymbol{x};Z\right)\label{eq:VP-subtraction}\\
\rho_{\kappa}^{\text{VP},1}\left(\boldsymbol{x};Z\right) & =\lim_{\delta\rightarrow0}\frac{Z}{\delta}\rho_{\kappa}^{\text{VP}}\left(\boldsymbol{x};\delta\right).\label{eq:VP-linear}
\end{align}

These equations assume that the $\cal{C}$-symmetry has been realized, and therefore, that the zero- and two-potential terms $\rho_{\kappa}^{\text{VP},0}$ and $\rho_{\kappa}^{\text{VP},2}$ vanish. At this point, the reader should be reminded that this subtraction eliminates the wanted physical Uehling contribution together with the unwanted non-physical logarithmic divergence. This should cause no worry since we know the exact expression for the Uehling potential that is given by \cite{Fullerton_Rinker_1976}
\begin{align}
-e\varphi_{\text{Ueh.}}\left(\boldsymbol{x}\right) & =-\frac{2\alpha\left(Z\alpha\right)}{3\pi}\hbar c\int d^{3}y\nonumber \\
 & \,\,\,\,\times\frac{\rho^{\text{n}}\left(\boldsymbol{x}-\boldsymbol{y}\right)}{\left|\boldsymbol{y}\right|}K_{1}\left(\frac{2\left|\boldsymbol{y}\right|}{\lambdabar}\right)\\
\text{with}\nonumber \\
K_{1}(x) & \ensuremath{=\int_{1}^{\infty}d\zeta e^{-x\zeta}\left(\frac{1}{\zeta^{2}}+\frac{1}{2\zeta^{4}}\right)\sqrt{\zeta^{2}-1}},
\end{align}
where $\lambdabar=\hbar/(mc)$ is the reduced Compton wavelength, and $\rho^{\text{n}}$ is an arbitrary nuclear distribution that enters Eq.(\ref{eq:external-potential}). This potential corrects the nuclear potential of Eq.(\ref{eq:external-potential}) at short distances and can be easily included as an effective potential in the Dirac equation, to account for the missing physics. Approximate expressions for this potential are provided by Wayne Fullerton and Rinker, in the last cited reference, in order to facilitate numerical evaluations. This physical (regularized) scalar potential solves the electrostatic Maxwell equation
\begin{equation}
\Delta\varphi_{\text{Ueh.}}\left(\boldsymbol{x}\right)=-\rho_{\text{Ueh.}}\left(\boldsymbol{x}\right)/\epsilon_{0},
\end{equation}
where $\rho_{\text{Ueh.}}$ is the regularized (and renormalized) version of the divergent one-potential VP density, given in Eq.(\ref{eq:VP-linear}) (summed over all values of $\kappa$). This density can be called the Uehling (VP) density; its expression is found in Ref. \cite[Eq.(45)]{Wichmann_Kroll1956}, for the point nuclei case.

Going back to the finite basis set problem, we note that if the set does not allow the realization of ${\cal C}$-symmetry, then VP densities that are of even orders of interaction with the external field (even orders in $Z$) shall not vanish, and will therefore corrupt the numerical result. In order to remove these unwanted terms, and therefore obtain cogent results, one can, instead of $\rho_{\kappa}^{\text{VP}}$ of Eq.(\ref{eq:VP-in-basis}), use the following VP density
\begin{equation}
\rho_{\kappa,{\cal C}}^{\text{VP}}\left(\boldsymbol{x};Z\right)=\frac{1}{2}\left[\rho_{\kappa}^{\text{VP}}\left(\boldsymbol{x};Z\right)-\rho_{\kappa}^{\text{VP}}\left(\boldsymbol{x};-Z\right)\right].\label{eq:VP-C-symmetric}
\end{equation}

This replacement forces the VP density to automatically obey ${\cal C}$-symmetry, even if the basis set, in which the density is constructed, does not do so. We now follow the previous reasoning and obtain the following expression for the many-potential VP density expression
\begin{equation}
\begin{aligned} & \rho_{\kappa,{\cal C}}^{\text{VP},n\geq3}\left(\boldsymbol{x};Z\right)\\
 & =\rho_{\kappa,{\cal C}}^{\text{VP}}\left(\boldsymbol{x};Z\right)-\lim_{\delta\rightarrow0}\frac{Z}{\delta}\rho_{\kappa,{\cal C}}^{\text{VP}}\left(\boldsymbol{x};\delta\right)\\
 & =\frac{1}{2}\left[\rho_{\kappa}^{\text{VP}}\left(\boldsymbol{x};Z\right)-\rho_{\kappa}^{\text{VP}}\left(\boldsymbol{x};-Z\right)\right]\\
 & -\lim_{\delta\rightarrow0}\frac{Z}{2\delta}\left[\rho_{\kappa}^{\text{VP}}\left(\boldsymbol{x};\delta\right)-\rho_{\kappa}^{\text{VP}}\left(\boldsymbol{x};-\delta\right)\right].
\end{aligned}\label{eq:VP-C-symmetric-non-linear}
\end{equation}

We note that in the case where the finite basis obeys the ${\cal C}$-symmetry, the initial VP density $\rho_{\kappa}^{\text{VP}}$ of Eq.(\ref{eq:VP-in-basis}) becomes equal to the new VP density $\rho_{\kappa,{\cal C}}^{\text{VP}}$ of Eq.(\ref{eq:VP-C-symmetric}). This last formula shall be used within the KB scheme where the ${\cal C}$-symmetry is generally violated.

\section{Numerical computations}

\label{sec:Numerical-computations}

In this section, we shall present some computational results of the VP density in the finite-basis approximation. The first part shall concern a qualitative improvement of the numerical results, driven by $\cal{C}$-symmetry, and in the second one, we shall see how quantitative results can be efficiently obtained. In the presented calculations, we have used the fine-structure constant value of $\alpha=1/137.036$ instead of the recommended value of $\alpha=1/137.035999084\left(21\right)$ by CODATA2018 \cite[Table XXXI]{2021codata2018}, allowing direct comparison of our VP density results, with the previous results of Mohr \textit{et al.} \cite[section 4.2]{mohr1998qed}. In the computation of the many-potential VP densities of Eqs.(\ref{eq:VP-subtraction},\ref{eq:VP-linear}) and Eq.(\ref{eq:VP-C-symmetric-non-linear}), we have chosen the small nuclear charge parameter to be $\delta=10^{-6}$. Since we are using the lowest-order forward finite-difference formula, this implies that the error associated with our derivatives is of order ${\cal{O}}(10^{-6})$. To reduce this error, smaller values of $\delta$ and/or higher-order finite-difference expressions can be employed. All presented results were computed using Wolfram Mathematica \cite{Mathematica}.

\subsection{Nuclear models}
\label{sec:Nuclear-models}
Our calculations include the following nuclear models
\begin{enumerate}
\item The point nucleus model, where the nuclear distribution and its associated scalar potential, entering Eq.(\ref{eq:external-potential}), are respectively given by
\begin{align}
\rho^{\text{n}}\left(\boldsymbol{x}\right) & =\delta\left(\boldsymbol{x}\right),\\
-e\phi^{\text{ext}.}\left(\boldsymbol{x}\right) & =-\frac{Z\alpha}{r}\hbar c.
\end{align}
\item The shell nucleus (hollow sphere) model, where we correspondingly have
\begin{align}
\rho^{\text{n}}\left(\boldsymbol{x}\right) & =\frac{1}{4\pi r_{\text{n}}^{2}}\delta\left(r-r_{\text{n}}\right)\\
-e\phi^{\text{ext}.}\left(\boldsymbol{x}\right) & =-\frac{Z\alpha}{r_{>}}\hbar c,
\end{align}
where $r_{>}=\max\left(r,r_{\text{n}}\right)$ and $r_{\text{n}}$ represents the shell radius after which the electric potential transitions from a constant function, to the point nucleus $1/r$ behavior. For the uranium atom, this parameter shall be set to $r_{\text{n}}=5.86\text{ fm}$, following Mohr \textit{et al.} \cite[section 4.2]{mohr1998qed}.
\end{enumerate}
In addition to the shell nucleus, standard extended nuclear models include Gaussian-, Fermi-, and ball-distributions (volumetric charge density), where each model is associated with the appropriate parameter(s). Details concerning these potentials can be found in Refs.  \cite{VISSCHER1997207,Andrae2000,nrsbook}. The choice of a nuclear model is usually made with respect to rendering the computation more practical. This reasoning is justified by the fact that the nuclear-size effect on the electron energy shift is dominated by a term that is proportional to the root mean square (RMS) charge radius $\small{\langle r^{2}\rangle}^{1/2}=(\smallint d^{3}xr^{2}\rho^{\text{n}}(\boldsymbol{x}))^{1/2}$; see for instance Refs. \cite[section 8.3]{froese1997computational} and \cite{almoukhalalati2016electron}. An empirical formula for this RMS function as a function of the atomic mass $A$ was provided by Johnson and Soff in Ref. \cite[Eq.(20)]{JOHNSON1985405}, allowing to determine nuclear parameters (such as $r_{\text{n}}$), as done by Visscher and Dyall in Ref. \cite{VISSCHER1997207}.

\subsection{Gaussian basis functions}
\label{subsec:Gaussian-basis-functions}

In our finite basis set calculations we shall use the following large 
and small Gaussian basis functions
\begin{align}
\pi_{\kappa,i}^{+}\left(r\right) & =r^{\left|\kappa+\frac{1}{2}\right|+\frac{1}{2}}e^{-\zeta_{\kappa,i}^{+}r^{2}} \label{eq:pi_+}\\
\pi_{\kappa,i}^{-}\left(r\right) & =r^{\left|\kappa-\frac{1}{2}\right|+\frac{1}{2}}e^{-\zeta_{\kappa,i}^{-}r^{2}}\label{eq:pi_-}.
\end{align}

For each radial problem associated with some $\kappa$ quantum number, we shall introduce a set of exponents $\zeta_{\kappa,i}^{\pm}$ for $i=1,\ldots,n_{\kappa}^{\pm}$, where the $\pm$ sign is added to distinguish between large and small component exponents. In addition, we note that the radial powers of these functions are chosen such that they describe the right leading order in $r$, of the exact solutions, at very short distances from the origin for both 1) the spherical free particle problem \cite[section 2.3.1]{Csymm_salman_saue}, as well as 2) the extended nucleus nuclear model case \cite[section 5.4.1]{Grant}. In addition, it was noted by Ishikawa \textit{et al.} \cite{ISHIKAWA1985130} that these Gaussian functions follow the exact next radial order(s). This indicates that Gaussian functions are well suited to mimic the radial functions behavior within the (finite) nuclear region. Furthermore, the mathematical importance of these functions comes from the fact that the radial integrals (in the matrix representation) can be analytically evaluated, therefore, avoiding numerical integrations. For future purposes, we note that in addition to these interesting features, Gaussian-type functions play an essential role in molecular calculations due to the Gaussian product rule that is associated with multi-center two-electron integrations, first noted by Boys in Ref. \cite{boys1950electronic}; see also Refs. \cite[appendix A]{szabo_Ostlund_2012modern} and \cite[section 9.2]{helgaker2014molecular}. It should be kept in mind that Gaussian functions have a faster decay rate than exact solutions (exponential decay); this should cause no problem since we aim to study the VP process, which is a very local effect. We shall use Gaussian exponents that are generated through the even-tempering prescription
\begin{equation}
\zeta_{\kappa,i}=\zeta_{\kappa,1}\left(\zeta_{\kappa,n}/\zeta_{\kappa,1}\right)^{\frac{i-1}{n-1}},\quad\text{for }i=1,\ldots,n
\end{equation}
where we shall specify the smallest and largest exponents $\zeta_{\kappa,1}$ and $\zeta_{\kappa,n}$, in addition to the number of exponents $n$. Throughout this work, we shall use three sets of Gaussian exponents whose associated parameters are tabulated in Table \ref{tab:Gaussian-bases-parameters.}.

\begin{table}[h]
\renewcommand{\arraystretch}{1.2}
\begin{ruledtabular}
\begin{tabular}{cccc}
Basis & $\zeta_{\kappa,1}$ & $\zeta_{\kappa,n}$ & $n$\\
\colrule 
\hspace*{0.18cm}10G & $10^{3}$ & $10^{7}$ & $10$\\
\hspace*{0.18cm}50G & $10^{3}$ & $10^{11}$ & $50$\\
150G & $10^{3}$ & $10^{11}$ & $150$\\
\end{tabular}
\caption{\label{tab:Gaussian-bases-parameters.}Gaussian bases parameters.}
\end{ruledtabular}
\end{table}

In all of the presented results, we shall set the same Gaussian exponent lists for both $\pm\kappa$ problems as well as both large and small component functions. This setting corresponds to $j$-bases, discussed in Sec. \ref{subsec:-symmetry-in-the-finite-basis}, and leads to the ${\cal C}$-symmetry realization in the DKB framework, as seen from Eq.(\ref{eq:zetas-C-DKB}), and indicated in our previous works of Refs. \cite[section 2.3.1]{Csymm_salman_saue} and \cite[section 2.11.6]{salman2022quantum}.

\subsection{Free-electron VP density}

\begin{figure}
\subfloat[\label{fig:Free-KB-calculation}KB calculation]{
\includegraphics[width= 8.6 cm]{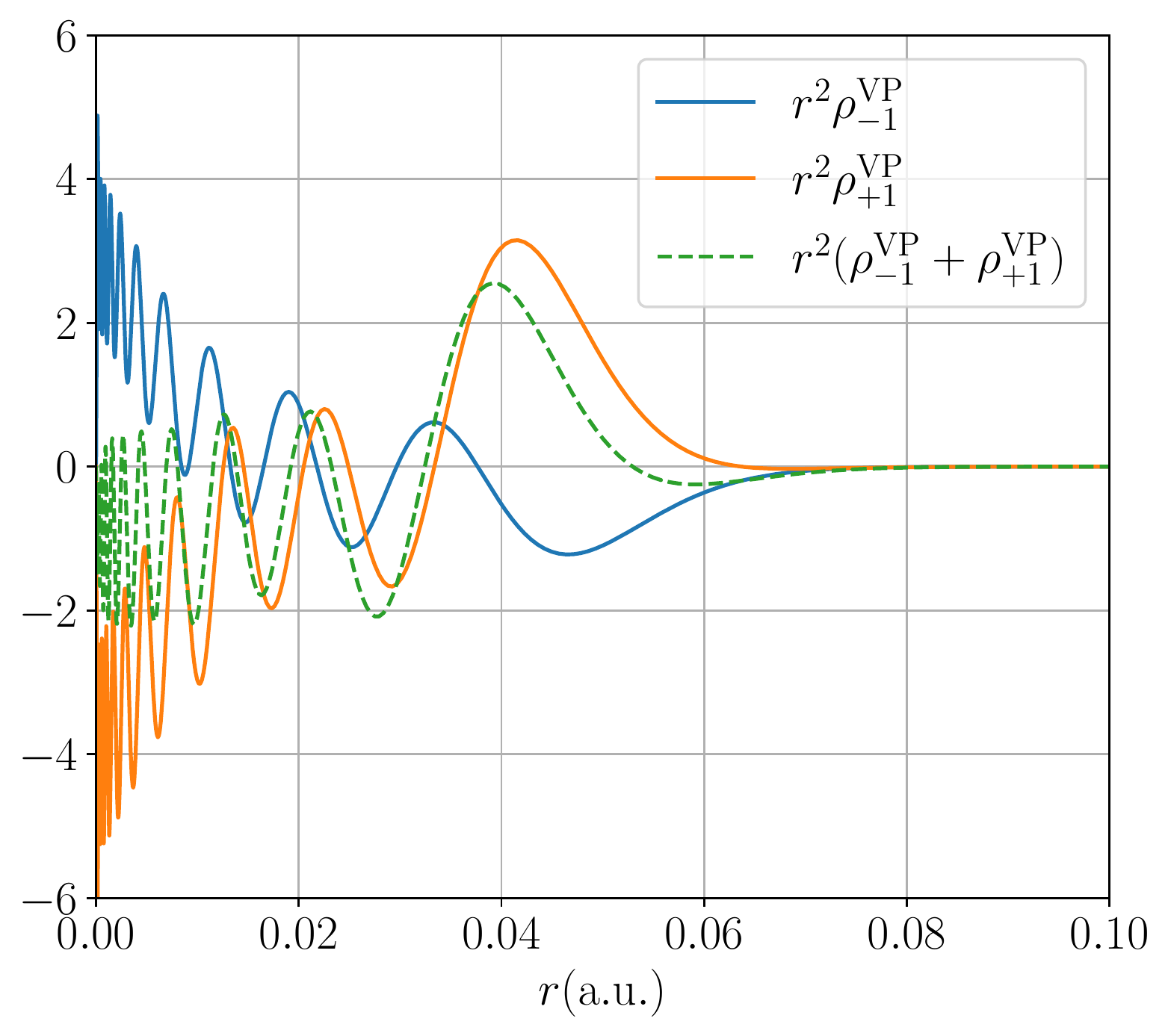}}\\
\subfloat[\label{fig:Free-DKB-calculation}DKB calculation]{
\includegraphics[width= 8.6 cm]{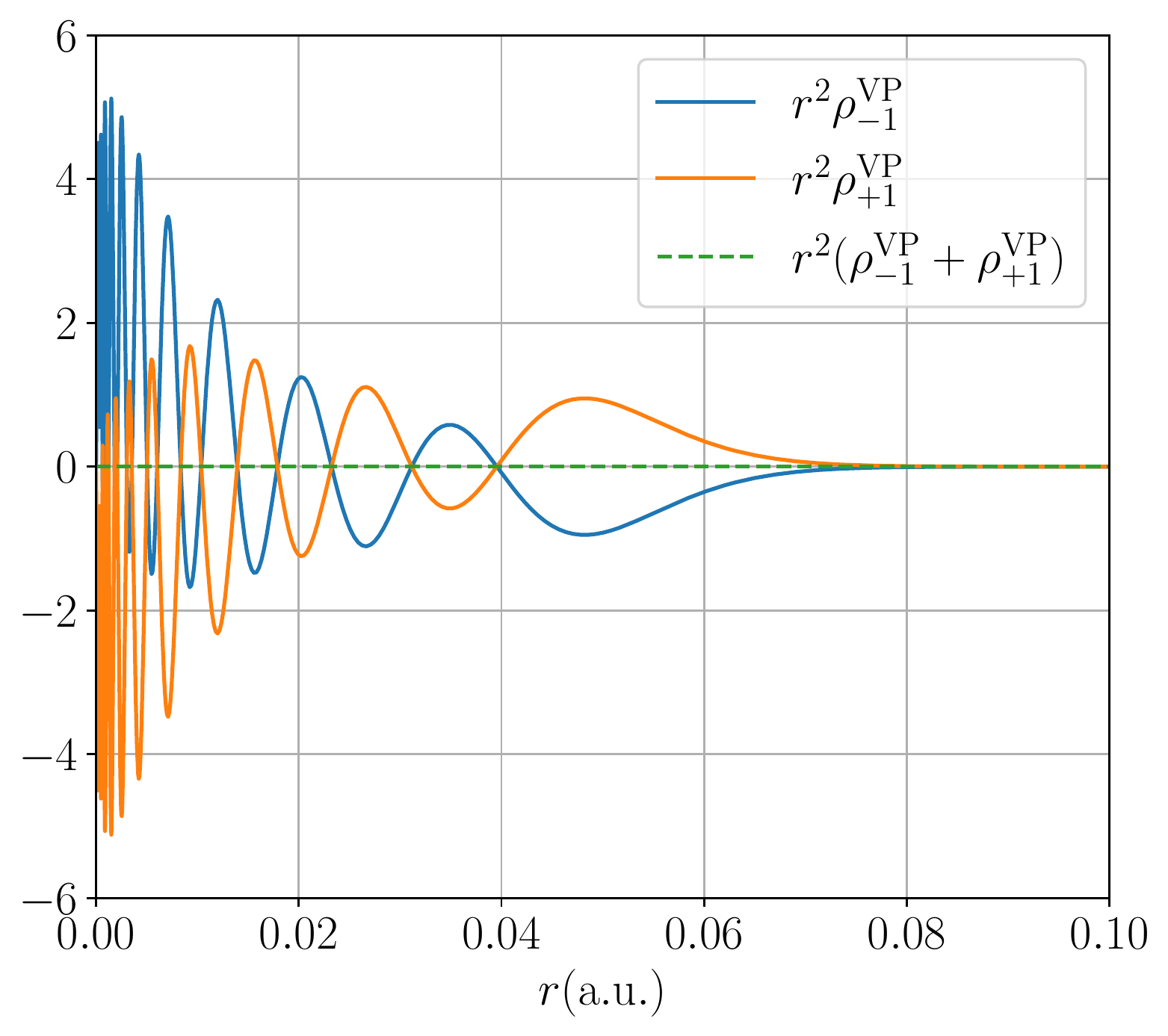}}
\caption{Free VP densities using 10G basis.}
\end{figure}

As discussed in sections \ref{sec:Theory}, in the free-particle spherical problem, the total VP density vanishes due to cancellation between solutions of opposite signs of energy and $\kappa$. We have therefore performed free-particle calculations ($Z=0$) of the $\kappa=\pm1$ problems using the 10G Gaussian basis of Table \ref{tab:Gaussian-bases-parameters.}. In the first calculation, we compute the total VP density of Eq.(\ref{eq:VP-in-basis}) in the KB basis construction, where the ${\cal C}$-symmetry is violated, for both $\kappa=\pm 1$ problems, and present the obtained results in Fig. \ref{fig:Free-KB-calculation}. This figure shows a non-vanishing sum of the two VP density components and indeed indicates ${\cal C}$-symmetry violation. In the second calculation, we computed the same VP density, within the DKB construction where the ${\cal C}$-symmetry is obeyed, and present the corresponding results in Fig. \ref{fig:Free-DKB-calculation}. Contrary to the previous result, we find a total cancellation (within numerical precision) between VP density components ($\pm \kappa$), as also noted by Grant and Quiney \cite{GrantQuiney2022atoms10040108}.

We now turn to the atomic problem, where $Z\neq0$, and show how ${\cal C}$-symmetry can guide us in obtaining more promising computational results.

\subsection{Total VP density}
\begin{figure}
\subfloat[\label{fig:KB-large-distances}KB calculation]{
\includegraphics[width= 8.6 cm]{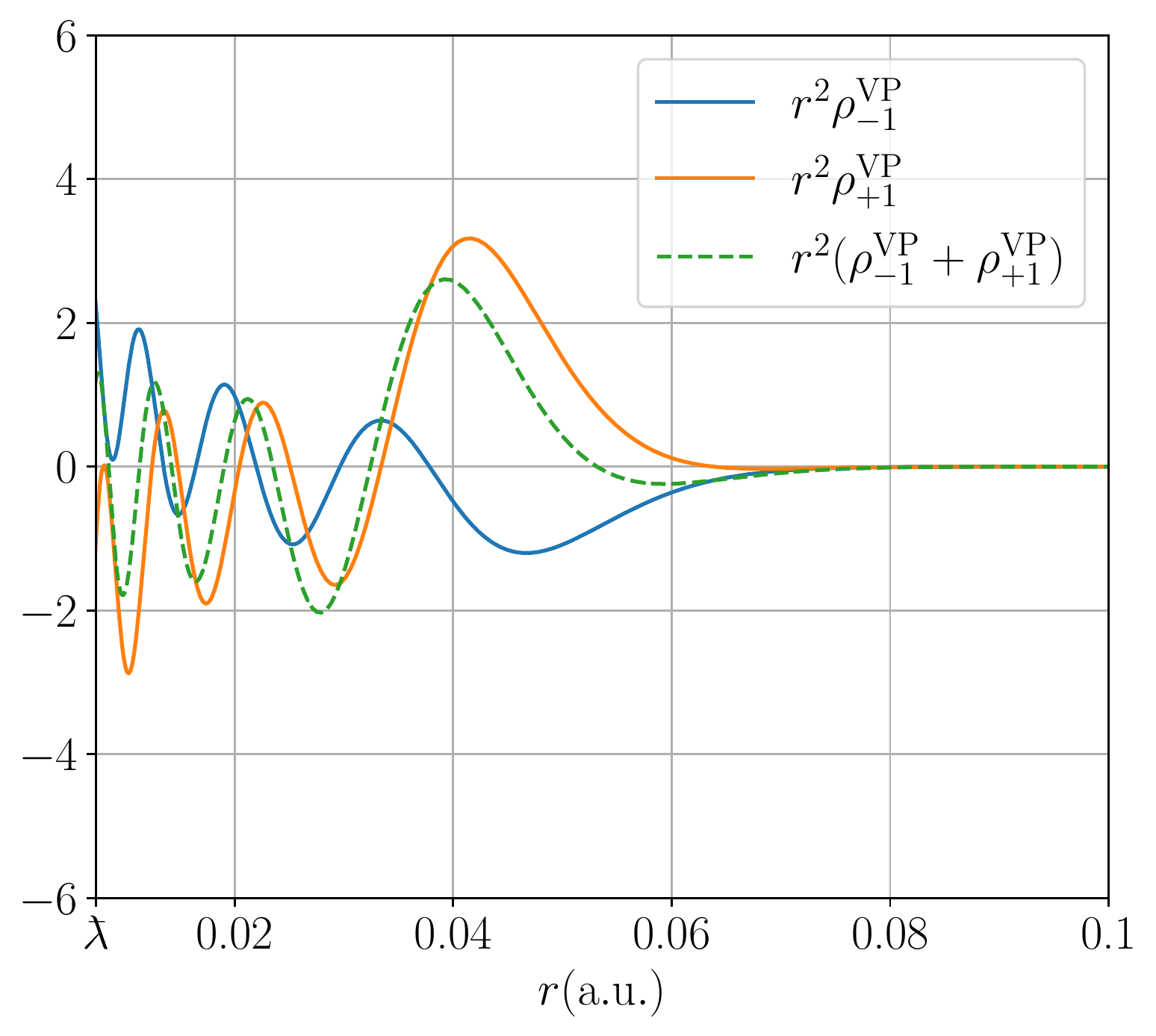}}\\
\subfloat[\label{fig:DKB-large-distances}DKB calculation]{
\includegraphics[width= 8.6 cm]{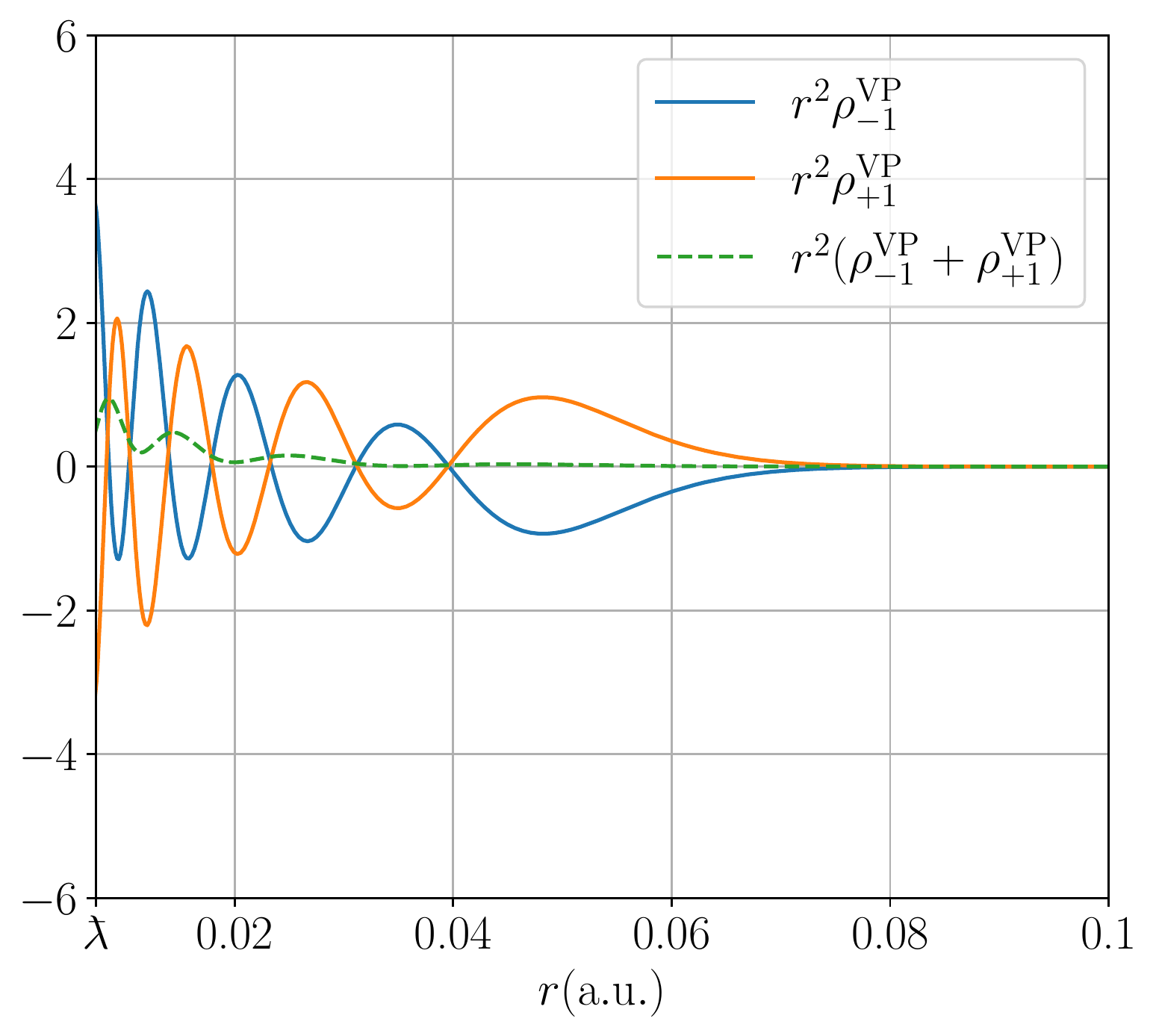}
}
\caption{\label{fig:VP-large-distances} VP density for the point nucleus
Uranium.}
\end{figure}

We run the same previous calculations (with KB and DKB constructions), but this time with a point nucleus of $Z=92$,  and take a closer look at the VP density behavior at distances $r\geq\lambdabar$. Results are presented in Fig. \ref{fig:VP-large-distances}. Again, contrary to the KB calculation, the DKB calculation provides better physical results by yielding a decaying VP polarization density at distances larger than the reduced Compton wavelength, as seen in Fig. \ref{fig:DKB-large-distances}, contrary to Fig. \ref{fig:KB-large-distances}, where a spurious contribution is still surviving. 

The total VP density can be expanded in powers of the external potential ($Z$), as given in Eq.(\ref{eq:VP-expansion}), and the first-order contribution comes from the free VP density $\rho_{\kappa}^{\text{VP}}\left(\boldsymbol{x};0\right)$. In the KB construction, where this contribution does not vanish, the total VP density gets contaminated by the non-vanishing free VP density, as seen when comparing Fig. (\ref{fig:Free-KB-calculation}) to  Fig. (\ref{fig:KB-large-distances}). Additional contaminations shall come from all non-vanishing (spurious) VP contributions of even orders in $Z$, with decreasing amplitudes.

Although any numerical evaluation (within the finite-basis approximation) of the VP density will yield a finite numerical result, this result is still divergent due to the linear contribution (in the nuclear charge parameter $Z$). This divergent behavior can be observed by gradually increasing the finite basis set size (including more Gaussian exponents) and realizing that the obtained density never converges, notably when more localized functions (with larger exponents) are appended in the finite basis. For this reason, we shall next consider removing the linear contribution through a simple procedure and isolate the many-potential VP density, that is free of divergences.

\subsection{Many-potential VP density with DKB}

Using the Green's function construction suggested by Wichmann and Kroll \cite{Wichmann_Kroll1956}, Soff and Mohr \cite{SoffMohr1988PhysRevA.38.5066} wrote the VP density expression in terms of the analytical expression of the Dirac Coulomb Green's function in the presence of a shell nucleus; they then subtracted the linear contribution that is written in terms of the shell nucleus potential and the free Green's function. A numerical evaluation of the residual integrations was performed, and the many-potential VP density for the first $\kappa=\pm1,\ldots,\pm5$ problems were presented in \cite[section 4.2, Fig. 9]{mohr1998qed}.

We, on the other hand, have decided to tackle the problem from the finite basis set perspective, and shown that we are capable to reproduce the same results up to a high degree of precision, at lower computational cost (no needed numerical integration), and using arbitrary radial nuclear charge distributions.

\begin{figure}
\begin{subfigure}{0.999\linewidth}
\includegraphics[width= 8.6 cm,trim={0.3cm 0.3cm 0.3cm 1.4cm},clip]{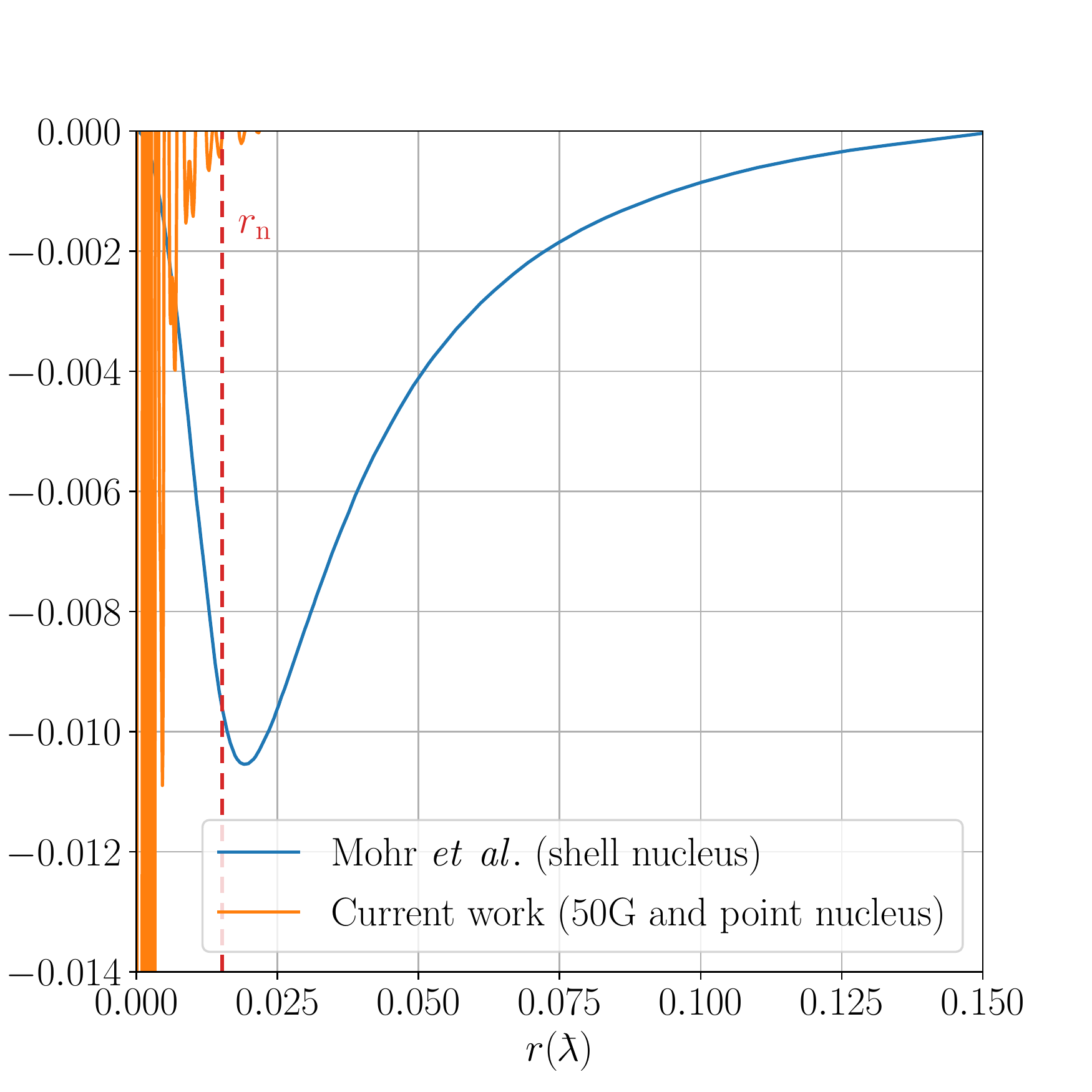}
\caption{\label{fig:Non-linear-DKB-Point-50G1}Short distances}
\end{subfigure}\\
\begin{subfigure}{0.999\linewidth}
\includegraphics[width= 8.6 cm,trim={0.3cm 0.3cm 0.3cm 1.4cm},clip]{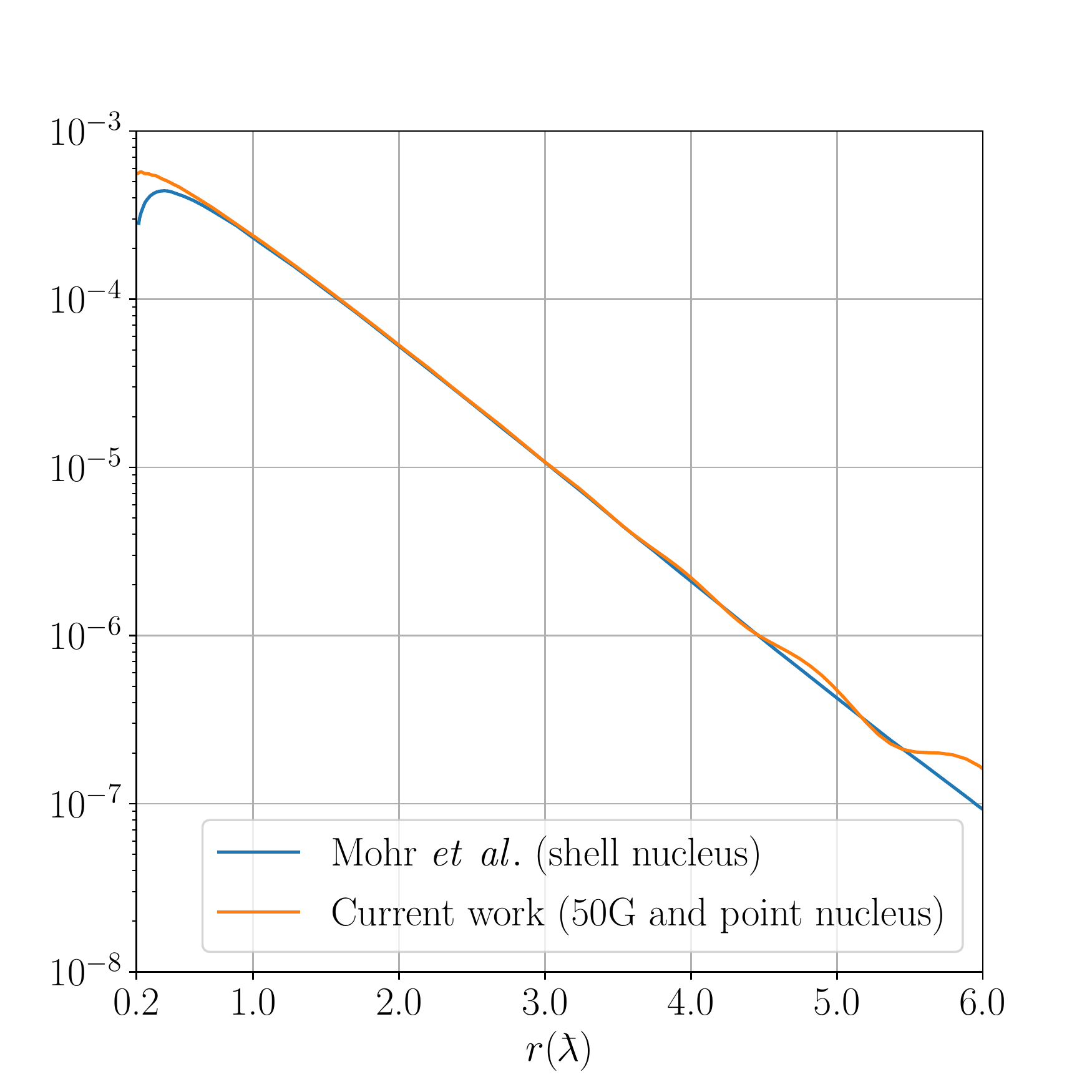}
\caption{\label{fig:Non-linear-DKB-Point-50G2}Large distances}
\end{subfigure}%
\caption{\label{fig:Non-linear-DKB-Point-50G}Many-potential VP density $\lambdabar r^2\rho_{|\kappa|=1}^{\text{VP},n\geq3}$ computed using DKB, the 50G basis, and point nucleus.}
\end{figure}

We have first evaluated the many-potential VP density for the one-electron uranium atom ($Z=92$) of a point nuclear distribution using the 50G basis and present the result in Fig. \ref{fig:Non-linear-DKB-Point-50G}. In the upper and lower panels, we plot the many-potential VP densities at short, and relatively large distances, respectively. The dashed red line is positioned at the nuclear radius $r_{\text{n}}$, discussed in Sec. \ref{sec:Nuclear-models}. We observe a very low-quality VP density near the point nucleus; this problem persists when the basis set size is gradually increased and can be traced back to the following two reasons. The first reason is that our radial solutions were constructed within the Gaussian basis, which does not describe the right radial behavior of the wavefunction near a point nucleus, as observed in Ref. \cite[chapter 7]{DyallFaegriRQC2007}. Secondly, the various kinetic balance constructions (KB, IKB, and DKB) assume that the nuclear potential obeys $\left|\phi^{\text{ext}}\left(r\right)\right|\ll mc^{2}$; this is clearly not the case of the point nucleus, notably in the limit $r\rightarrow0$. Nevertheless, we observe that our point nucleus results are able to reproduce the finite nucleus result of Mohr \textit{et al.} \cite{mohr1998qed} at distances $r>\lambdabar$. Basis sets that are designed to describe radial Dirac wavefunctions of the point nucleus problem (L- and S-spinors), and account for its singularities, are discussed in detail by Grant \cite[sections 5.8 and 5.9]{Grant}. We finally note that additional calculations show that with a larger basis set size, the wiggly behavior we have at $r>4\lambdabar$ in Fig. \ref{fig:Non-linear-DKB-Point-50G2} can be totally damped.

\begin{figure}
\begin{subfigure}{0.999\linewidth}
\includegraphics[width= 8.6 cm,trim={0.3cm 0.3cm 0.3cm 1.4cm},clip]{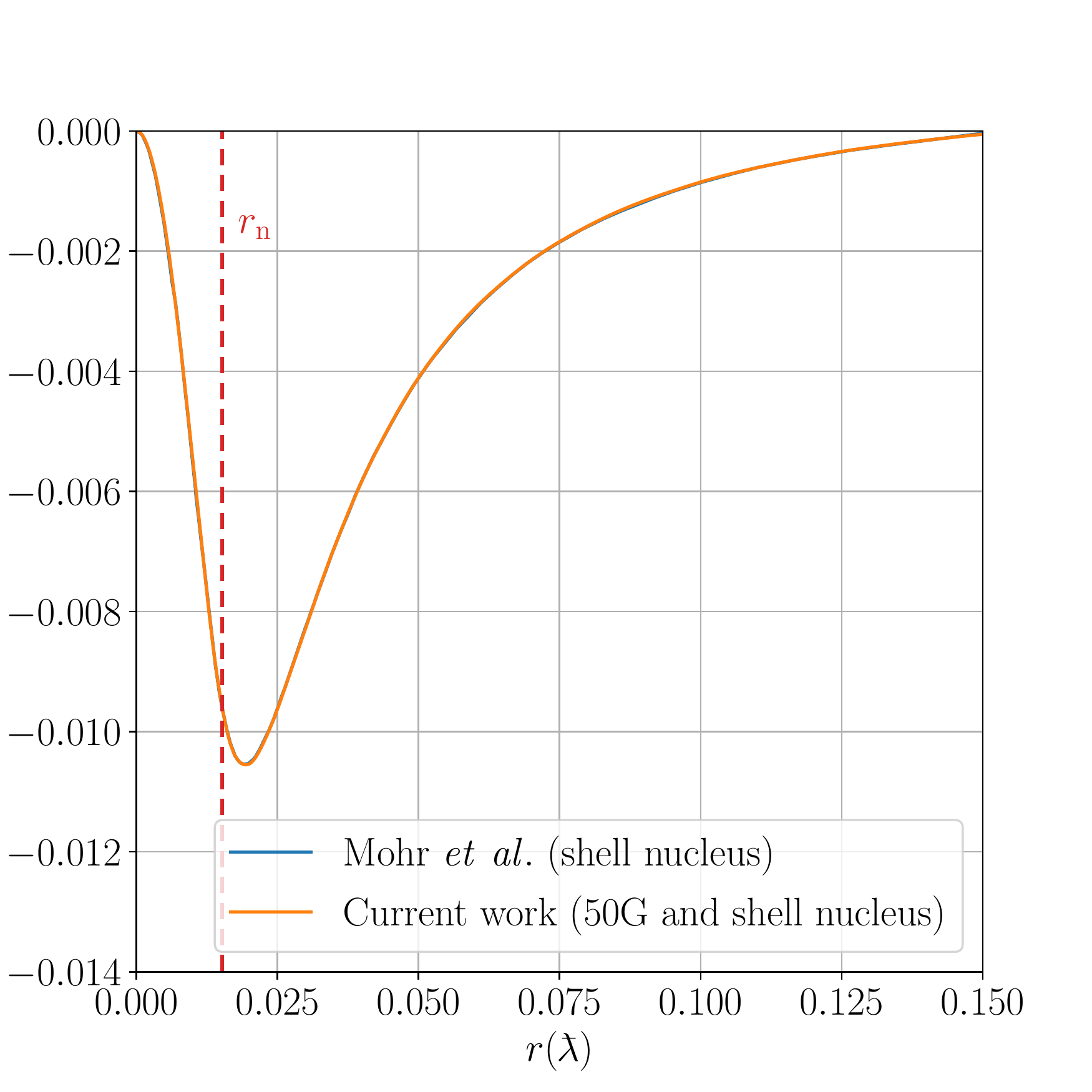}
\caption{Short distances}
\end{subfigure}\\
\begin{subfigure}{0.999\linewidth}
\includegraphics[width= 8.6 cm,trim={0.3cm 0.3cm 0.3cm 1.4cm},clip]{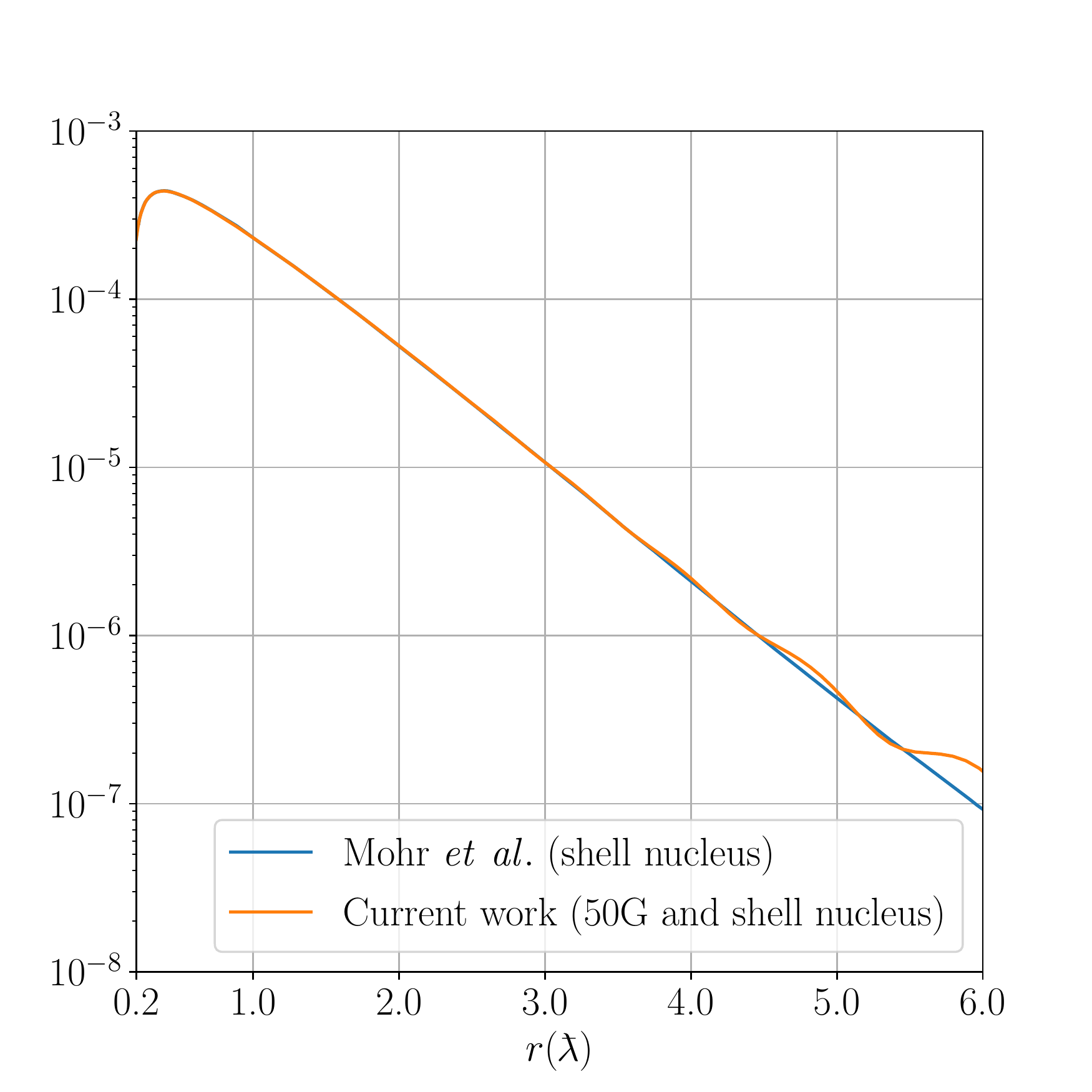}
\caption{Large distances}
\end{subfigure}%
\caption{\label{fig:Non-linear-DKB-Shell-50G}Many-potential VP density $\lambdabar r^2\rho_{|\kappa|=1}^{\text{VP},n\geq3}$ computed using DKB, the 50G basis, and shell nucleus.}
\end{figure}

We have next performed the same calculation but this time using a shell nucleus model of radius $r_{\text{n}}$ and present the obtained result in Fig. \ref{fig:Non-linear-DKB-Shell-50G}. We clearly observe a large agreement with the results of Mohr \textit{et al.} at both small and (relatively) large distances.

\begin{figure}
\begin{subfigure}{0.999\linewidth}
\includegraphics[width= 8.6 cm,trim={0.3cm 0.3cm 0.3cm 1.4cm},clip]{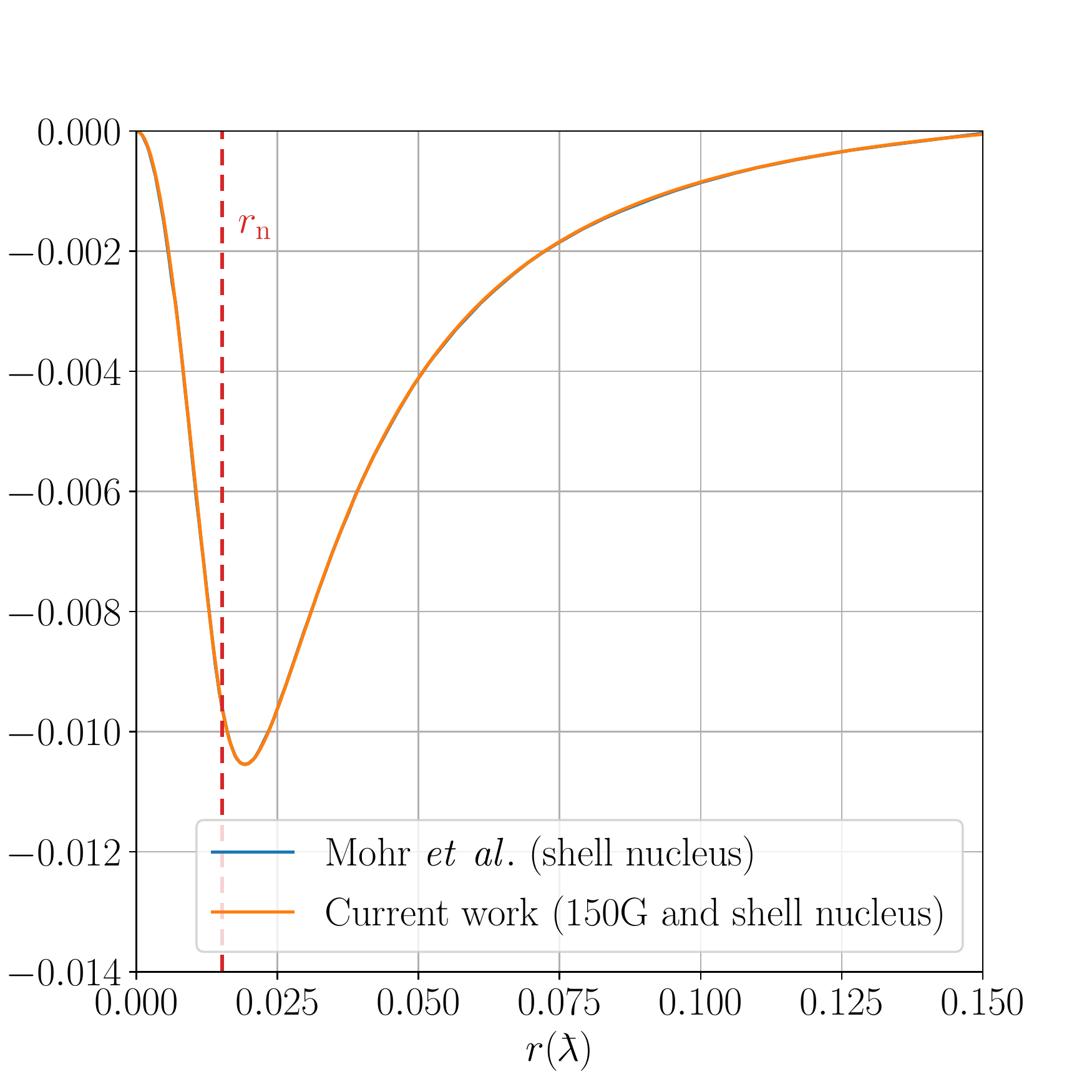}
\caption{Short distances}
\end{subfigure}\\
\begin{subfigure}{0.999\linewidth}
\includegraphics[width= 8.6 cm,trim={0.3cm 0.3cm 0.3cm 1.4cm},clip]{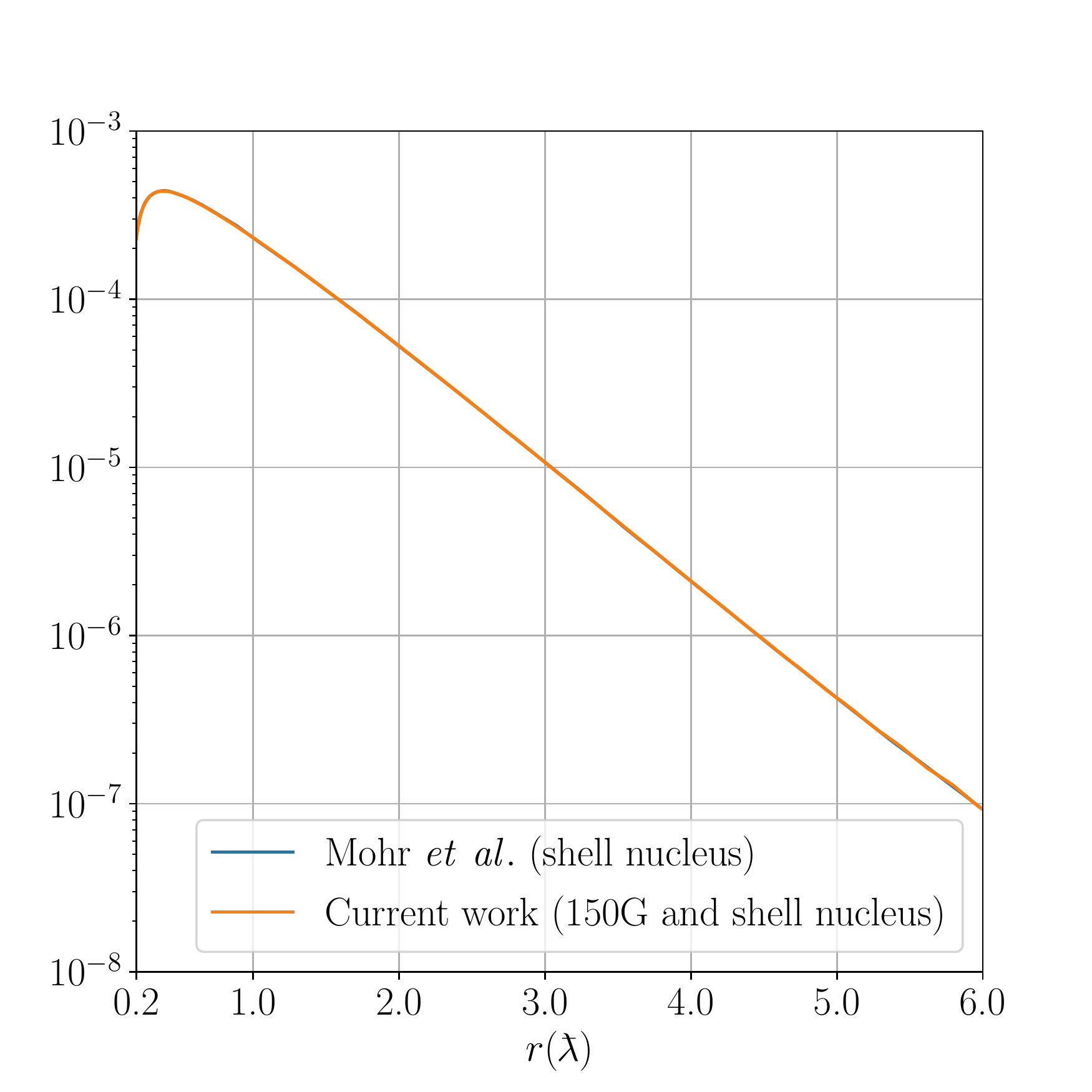}
\caption{Large distances}
\end{subfigure}%
\caption{\label{fig:Non-linear-DKB-Shell-150G}Many-potential VP density $\lambdabar r^2\rho_{|\kappa|=1}^{\text{VP},n\geq3}$ computed using DKB, the 150G basis, and shell nucleus.}
\end{figure}

Repeating the last calculation with a larger (150G) basis set size yields a perfect agreement, notably at $r>4\lambdabar$, as presented in Fig. \ref{fig:Non-linear-DKB-Shell-150G}.%

\subsection{Many-potential VP density with KB}

We have shown that using the KB prescription in constructing relativistic basis sets, one obtains non-physical results such as a non-vanishing free VP density and a non-decaying atomic VP density at distances larger than the reduced Compton wavelength $\lambdabar$. 

In this section, we shall show that by employing our formulas discussed in Sec. \ref{subsec:VP-in-the-finite-basis}, the ${\cal C}$-symmetry gets automatically obeyed and one can surpass the spurious limitations associated with the KB prescription, or any other basis construction that violates ${\cal C}$-symmetry. In addition, we shall show that an efficient evaluation of the many-potential VP density within the KB construction is possible. 

\begin{figure}
\begin{subfigure}{0.999\linewidth}
\includegraphics[width= 8.6 cm,trim={0.3cm 0.3cm 0.3cm 1.4cm},clip]{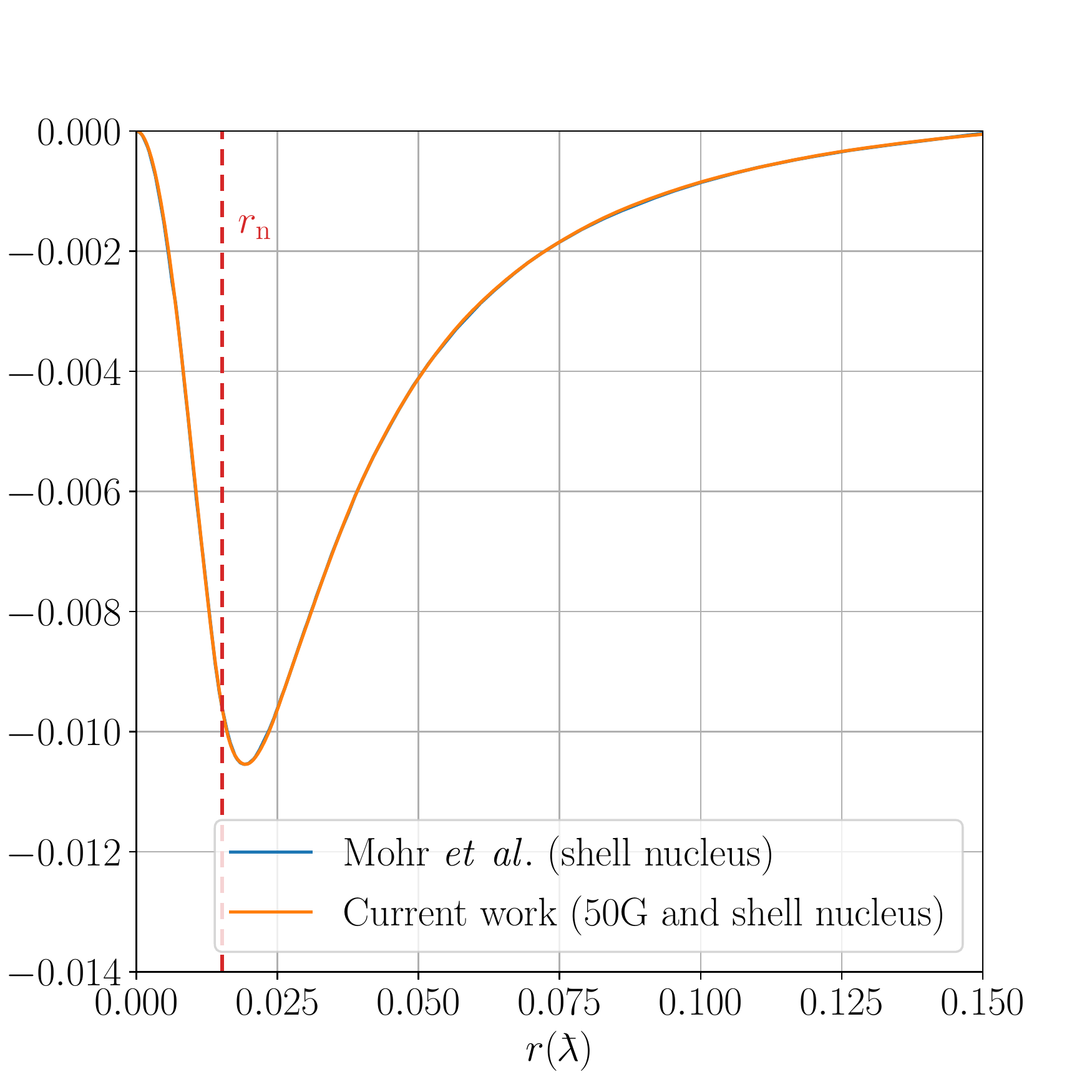}
\caption{Short distances}
\end{subfigure}\\
\begin{subfigure}{0.999\linewidth}
\includegraphics[width= 8.6 cm,trim={0.3cm 0.3cm 0.3cm 1.4cm},clip]{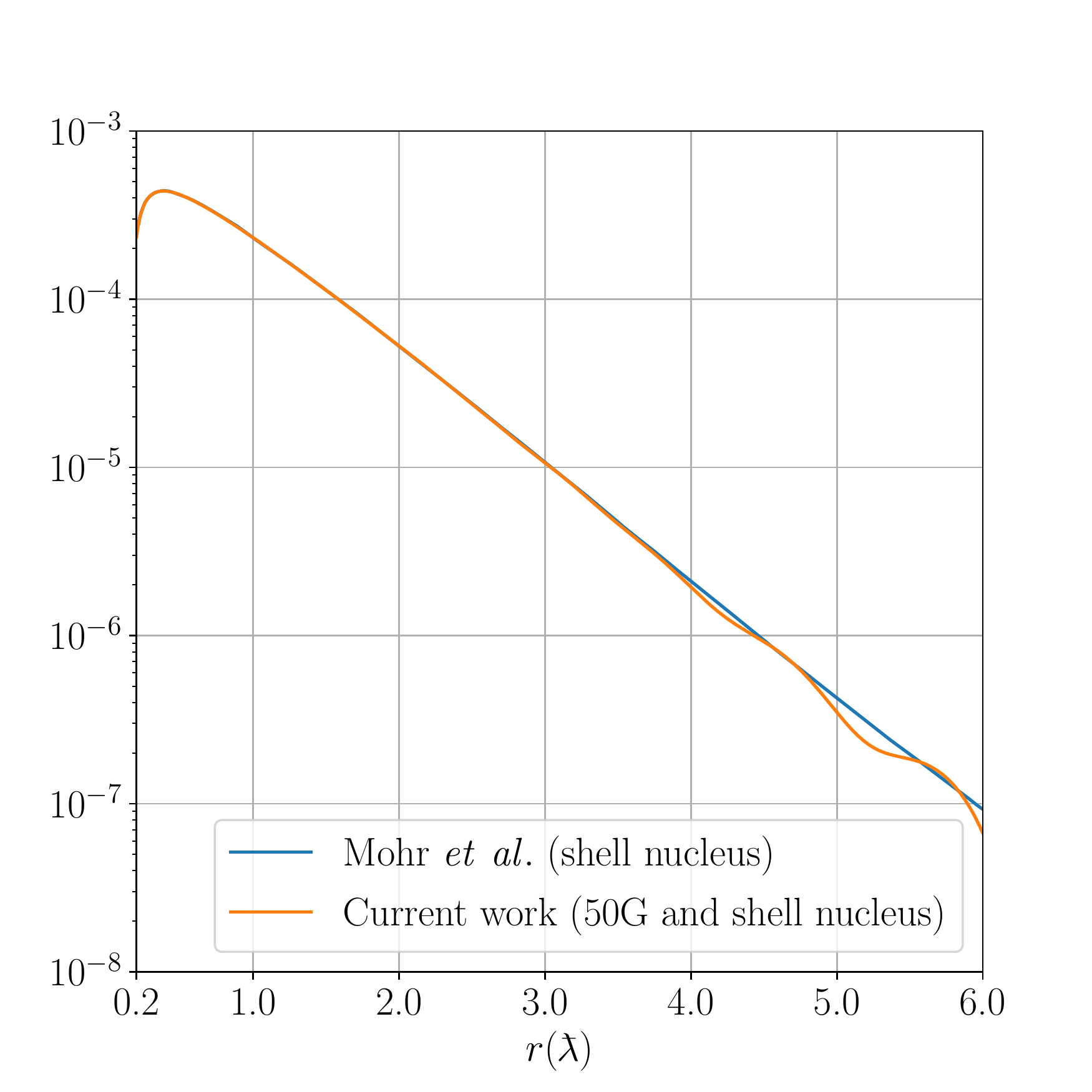}
\caption{Large distances}
\end{subfigure}%
\caption{\label{fig:Non-linear-RKB-50G}Many-potential VP density $\lambdabar r^2\rho_{|\kappa|=1}^{\text{VP},n\geq3}$ computed using KB, the 50G basis, and shell nucleus.}
\end{figure}

We remind the reader that in the free particle case ($Z=0$), both total and many-potential VP densities of Eq.(\ref{eq:VP-C-symmetric}) and Eq.(\ref{eq:VP-C-symmetric-non-linear}), vanish. To demonstrate the usefulness of Eq.(\ref{eq:VP-C-symmetric-non-linear}), we employ it in computing the many-potential VP density for the one-electron uranium problem and use solutions that are calculated within the KB basis construction. We ran four calculations, with $Z=\pm\delta$, and $\pm92$, on the shell nucleus problem, and present the final VP density in Fig. \ref{fig:Non-linear-RKB-50G}. The obtained results agree very well with the ones of Mohr \textit{et al.} and prove that the many-potential VP density can be efficiently and accurately computed in standard molecular programs (finite basis) that are typically based on the KB construction (where the $\cal{C}$-symmetry is generally violated).

\section{Conclusion}

In this paper, we have investigated the construction of the VP charge density for one-electron atoms within the finite-basis approximation, with a particular focus on the many-potential contribution that is free of divergences. In addition, we have shown that in our case the VP three-current vanishes due to time-reversal symmetry.

Concerning the finite basis problem, we have found that compliance with ${\cal C}$-symmetry is crucial to obtain physically valid VP density results. We note that within the DKB construction, the ${\cal C}$-symmetry realization is manifested by a vanishing total VP density once the proper matching between large and small basis functions is settled. Furthermore, we have computed this total VP density in our ${\cal C}$-symmetric basis, using an extended nuclear distribution (shell model), and subtracted out the linear contribution (in $Z$), where the logarithmic VP divergence is buried. The obtained results are found to be in excellent agreement with the former results of Mohr \textit{et al.} \cite[section 4.2]{mohr1998qed}. 

In the standard KB construction, the ${\cal C}$-symmetry is generally violated, and as a consequence, the computed VP density is found to be contaminated by spurious (non-vanishing) contributions. Moreover, we have shown that within this construction, the ${\cal C}$-symmetry can be forced by choosing the large component free-particle solution (spherical Bessel functions) to be the large component basis function. Due to the KB coupling, the small component basis function automatically becomes the small free-particle solution. The same analysis applies to the IKB construction. The main drawback of this basis is that it does not allow writing the radial integrals of the Dirac equation matrix representation in closed analytical forms; one is therefore obliged to employ numerical integration techniques. Furthermore, we have shown that the KB inadequacy in computing the VP density can be surpassed by imposing the ${\cal C}$-symmetry on the VP density expression instead of the basis set. This result indicates that any relativistic finite-basis (molecular or atomic) program whose Dirac matrix representation is constructed according to the KB prescription (as is the case of most molecular codes) can efficiently compute the many-potential VP effects; this approach transcends including the limited effective VP potential that is associated with the third-order VP correction to the Coulomb potential (of a point nucleus). 

We finally note that, besides numerical efficiency, the importance of our proposed many-potential VP density computation machinery lies in the fact that it can be applied to radial Dirac problems with arbitrary radial nuclear charge distributions. This method is, therefore, of particular significance for Dirac problems where analytical expressions of the associated Green's function are not at hand.

\begin{acknowledgments}
We thank Peter Mohr (NIST) for providing a higher-resolution version of the many-potential VP density of the one-electron uranium atom that was published in Ref. \cite[Fig. 9]{mohr1998qed}. We also thank Julien Toulouse (Paris) for helpful discussions. This project was funded by the European Research Council (ERC) under the European Union's Horizon 2020 research and innovation programme (grant agreement ID:101019907).
\end{acknowledgments}

\bibliographystyle{apsrev4-2.bst}
\bibliography{article}

\begin{thebibliography}{69}%
\makeatletter
\providecommand \@ifxundefined [1]{%
 \@ifx{#1\undefined}
}%
\providecommand \@ifnum [1]{%
 \ifnum #1\expandafter \@firstoftwo
 \else \expandafter \@secondoftwo
 \fi
}%
\providecommand \@ifx [1]{%
 \ifx #1\expandafter \@firstoftwo
 \else \expandafter \@secondoftwo
 \fi
}%
\providecommand \natexlab [1]{#1}%
\providecommand \enquote  [1]{``#1''}%
\providecommand \bibnamefont  [1]{#1}%
\providecommand \bibfnamefont [1]{#1}%
\providecommand \citenamefont [1]{#1}%
\providecommand \href@noop [0]{\@secondoftwo}%
\providecommand \href [0]{\begingroup \@sanitize@url \@href}%
\providecommand \@href[1]{\@@startlink{#1}\@@href}%
\providecommand \@@href[1]{\endgroup#1\@@endlink}%
\providecommand \@sanitize@url [0]{\catcode `\\12\catcode `\$12\catcode
  `\&12\catcode `\#12\catcode `\^12\catcode `\_12\catcode `\%12\relax}%
\providecommand \@@startlink[1]{}%
\providecommand \@@endlink[0]{}%
\providecommand \url  [0]{\begingroup\@sanitize@url \@url }%
\providecommand \@url [1]{\endgroup\@href {#1}{\urlprefix }}%
\providecommand \urlprefix  [0]{URL }%
\providecommand \Eprint [0]{\href }%
\providecommand \doibase [0]{https://doi.org/}%
\providecommand \selectlanguage [0]{\@gobble}%
\providecommand \bibinfo  [0]{\@secondoftwo}%
\providecommand \bibfield  [0]{\@secondoftwo}%
\providecommand \translation [1]{[#1]}%
\providecommand \BibitemOpen [0]{}%
\providecommand \bibitemStop [0]{}%
\providecommand \bibitemNoStop [0]{.\EOS\space}%
\providecommand \EOS [0]{\spacefactor3000\relax}%
\providecommand \BibitemShut  [1]{\csname bibitem#1\endcsname}%
\let\auto@bib@innerbib\@empty
\bibitem [{\citenamefont {Wichmann}\ and\ \citenamefont
  {Kroll}(1956)}]{Wichmann_Kroll1956}%
  \BibitemOpen
  \bibfield  {author} {\bibinfo {author} {\bibfnamefont {E.~H.}\ \bibnamefont
  {Wichmann}}\ and\ \bibinfo {author} {\bibfnamefont {N.~M.}\ \bibnamefont
  {Kroll}},\ }\href {https://doi.org/10.1103/PhysRev.101.843} {\bibfield
  {journal} {\bibinfo  {journal} {Phys. Rev.}\ }\textbf {\bibinfo {volume}
  {101}},\ \bibinfo {pages} {843} (\bibinfo {year} {1956})}\BibitemShut
  {NoStop}%
\bibitem [{\citenamefont {Hylton}(1984)}]{hylton1984reduced}%
  \BibitemOpen
  \bibfield  {author} {\bibinfo {author} {\bibfnamefont {D.}~\bibnamefont
  {Hylton}},\ }\href {https://doi.org/10.1063/1.526255} {\bibfield  {journal}
  {\bibinfo  {journal} {Journal of mathematical physics}\ }\textbf {\bibinfo
  {volume} {25}},\ \bibinfo {pages} {1125} (\bibinfo {year}
  {1984})}\BibitemShut {NoStop}%
\bibitem [{\citenamefont {Yerokhin}\ and\ \citenamefont
  {Maiorova}(2020)}]{Yerokhin-Maiorova2022-Green-function}%
  \BibitemOpen
  \bibfield  {author} {\bibinfo {author} {\bibfnamefont {V.~A.}\ \bibnamefont
  {Yerokhin}}\ and\ \bibinfo {author} {\bibfnamefont {A.~V.}\ \bibnamefont
  {Maiorova}},\ }\bibfield  {journal} {\bibinfo  {journal} {Symmetry}\ }\textbf
  {\bibinfo {volume} {12}},\ \href {https://doi.org/10.3390/sym12050800}
  {10.3390/sym12050800} (\bibinfo {year} {2020})\BibitemShut {NoStop}%
\bibitem [{\citenamefont {Grant}(2007)}]{Grant}%
  \BibitemOpen
  \bibfield  {author} {\bibinfo {author} {\bibfnamefont {I.~P.}\ \bibnamefont
  {Grant}},\ }\href {https://doi.org/10.1007/978-0-387-35069-1} {\emph
  {\bibinfo {title} {{Relativistic Quantum Theory of Atoms and Molecules}}}}\
  (\bibinfo  {publisher} {Springer-Verlag New York},\ \bibinfo {address} {New
  York},\ \bibinfo {year} {2007})\BibitemShut {NoStop}%
\bibitem [{\citenamefont {Swainson}\ and\ \citenamefont
  {Drake}(1991)}]{Swainson_Drake1991}%
  \BibitemOpen
  \bibfield  {author} {\bibinfo {author} {\bibfnamefont {R.~A.}\ \bibnamefont
  {Swainson}}\ and\ \bibinfo {author} {\bibfnamefont {G.~W.~F.}\ \bibnamefont
  {Drake}},\ }\href {https://doi.org/10.1088/0305-4470/24/1/020} {\bibfield
  {journal} {\bibinfo  {journal} {Journal of Physics A: Mathematical and
  General}\ }\textbf {\bibinfo {volume} {24}},\ \bibinfo {pages} {95} (\bibinfo
  {year} {1991})}\BibitemShut {NoStop}%
\bibitem [{\citenamefont {Hill}(2006)}]{Hill2006}%
  \BibitemOpen
  \bibfield  {author} {\bibinfo {author} {\bibfnamefont {R.}~\bibnamefont
  {Hill}},\ }\bibinfo {title} {{Hydrogenic Wave Functions}},\ in\ \href
  {https://doi.org/10.1007/978-0-387-26308-3_9} {\emph {\bibinfo {booktitle}
  {Springer Handbook of Atomic, Molecular, and Optical Physics}}},\ \bibinfo
  {editor} {edited by\ \bibinfo {editor} {\bibfnamefont {G.}~\bibnamefont
  {Drake}}}\ (\bibinfo  {publisher} {Springer New York},\ \bibinfo {address}
  {New York, NY},\ \bibinfo {year} {2006})\ pp.\ \bibinfo {pages}
  {153--171}\BibitemShut {NoStop}%
\bibitem [{\citenamefont {Uehling}(1935)}]{Uehling1935}%
  \BibitemOpen
  \bibfield  {author} {\bibinfo {author} {\bibfnamefont {E.~A.}\ \bibnamefont
  {Uehling}},\ }\href {https://doi.org/10.1103/PhysRev.48.55} {\bibfield
  {journal} {\bibinfo  {journal} {Phys. Rev.}\ }\textbf {\bibinfo {volume}
  {48}},\ \bibinfo {pages} {55} (\bibinfo {year} {1935})}\BibitemShut {NoStop}%
\bibitem [{\citenamefont {Blomqvist}(1972)}]{BLOMQVIST197295}%
  \BibitemOpen
  \bibfield  {author} {\bibinfo {author} {\bibfnamefont {J.}~\bibnamefont
  {Blomqvist}},\ }\href {https://doi.org/10.1016/0550-3213(72)90051-X}
  {\bibfield  {journal} {\bibinfo  {journal} {Nuclear Physics B}\ }\textbf
  {\bibinfo {volume} {48}},\ \bibinfo {pages} {95} (\bibinfo {year}
  {1972})}\BibitemShut {NoStop}%
\bibitem [{\citenamefont {Rinker}\ and\ \citenamefont
  {Wilets}(1973)}]{RinkerWilets1973PhysRevLett.31.1559}%
  \BibitemOpen
  \bibfield  {author} {\bibinfo {author} {\bibfnamefont {G.~A.}\ \bibnamefont
  {Rinker}}\ and\ \bibinfo {author} {\bibfnamefont {L.}~\bibnamefont
  {Wilets}},\ }\href {https://doi.org/10.1103/PhysRevLett.31.1559} {\bibfield
  {journal} {\bibinfo  {journal} {Phys. Rev. Lett.}\ }\textbf {\bibinfo
  {volume} {31}},\ \bibinfo {pages} {1559} (\bibinfo {year}
  {1973})}\BibitemShut {NoStop}%
\bibitem [{\citenamefont {Gyulassy}(1974)}]{Gyulassy_PhD_1974}%
  \BibitemOpen
  \bibfield  {author} {\bibinfo {author} {\bibfnamefont {M.}~\bibnamefont
  {Gyulassy}},\ }\emph {\bibinfo {title} {{Higher Order Vacuum Polarization for
  Finite Radius Nuclei: Application to Muonic Lead and Heavy Ion
  Collisions}}},\ \href {https://escholarship.org/uc/item/44t9359n} {Ph.D.
  thesis},\ \bibinfo  {school} {University of California, Berkeley.} (\bibinfo
  {year} {1974})\BibitemShut {NoStop}%
\bibitem [{\citenamefont {Rinker}\ and\ \citenamefont
  {Wilets}(1975)}]{Rinker_Wilets1975PhysRevA.12.748}%
  \BibitemOpen
  \bibfield  {author} {\bibinfo {author} {\bibfnamefont {G.~A.}\ \bibnamefont
  {Rinker}}\ and\ \bibinfo {author} {\bibfnamefont {L.}~\bibnamefont
  {Wilets}},\ }\href {https://doi.org/10.1103/PhysRevA.12.748} {\bibfield
  {journal} {\bibinfo  {journal} {Phys. Rev. A}\ }\textbf {\bibinfo {volume}
  {12}},\ \bibinfo {pages} {748} (\bibinfo {year} {1975})}\BibitemShut
  {NoStop}%
\bibitem [{\citenamefont {Borie}\ and\ \citenamefont
  {Rinker}(1982)}]{Borie_Rinker1982RevModPhys.54.67}%
  \BibitemOpen
  \bibfield  {author} {\bibinfo {author} {\bibfnamefont {E.}~\bibnamefont
  {Borie}}\ and\ \bibinfo {author} {\bibfnamefont {G.~A.}\ \bibnamefont
  {Rinker}},\ }\href {https://doi.org/10.1103/RevModPhys.54.67} {\bibfield
  {journal} {\bibinfo  {journal} {Rev. Mod. Phys.}\ }\textbf {\bibinfo {volume}
  {54}},\ \bibinfo {pages} {67} (\bibinfo {year} {1982})}\BibitemShut {NoStop}%
\bibitem [{\citenamefont {Soff}\ and\ \citenamefont
  {Mohr}(1988)}]{SoffMohr1988PhysRevA.38.5066}%
  \BibitemOpen
  \bibfield  {author} {\bibinfo {author} {\bibfnamefont {G.}~\bibnamefont
  {Soff}}\ and\ \bibinfo {author} {\bibfnamefont {P.~J.}\ \bibnamefont
  {Mohr}},\ }\href {https://doi.org/10.1103/PhysRevA.38.5066} {\bibfield
  {journal} {\bibinfo  {journal} {Phys. Rev. A}\ }\textbf {\bibinfo {volume}
  {38}},\ \bibinfo {pages} {5066} (\bibinfo {year} {1988})}\BibitemShut
  {NoStop}%
\bibitem [{\citenamefont {Brown}\ \emph {et~al.}(1975)\citenamefont {Brown},
  \citenamefont {Cahn},\ and\ \citenamefont
  {McLerran}}]{BrownCahnMcLerran1975PhysRevD.12.609}%
  \BibitemOpen
  \bibfield  {author} {\bibinfo {author} {\bibfnamefont {L.~S.}\ \bibnamefont
  {Brown}}, \bibinfo {author} {\bibfnamefont {R.~N.}\ \bibnamefont {Cahn}},\
  and\ \bibinfo {author} {\bibfnamefont {L.~D.}\ \bibnamefont {McLerran}},\
  }\href {https://doi.org/10.1103/PhysRevD.12.609} {\bibfield  {journal}
  {\bibinfo  {journal} {Phys. Rev. D}\ }\textbf {\bibinfo {volume} {12}},\
  \bibinfo {pages} {609} (\bibinfo {year} {1975})}\BibitemShut {NoStop}%
\bibitem [{\citenamefont {Neghabian}(1983)}]{Neghabian1983_PhysRevA.27.2311}%
  \BibitemOpen
  \bibfield  {author} {\bibinfo {author} {\bibfnamefont {A.~R.}\ \bibnamefont
  {Neghabian}},\ }\href {https://doi.org/10.1103/PhysRevA.27.2311} {\bibfield
  {journal} {\bibinfo  {journal} {Phys. Rev. A}\ }\textbf {\bibinfo {volume}
  {27}},\ \bibinfo {pages} {2311} (\bibinfo {year} {1983})}\BibitemShut
  {NoStop}%
\bibitem [{\citenamefont {Schmidt}\ \emph {et~al.}(1989)\citenamefont
  {Schmidt}, \citenamefont {Soff},\ and\ \citenamefont
  {Mohr}}]{SchmidtSoffMohr1989PhysRevA.40.2176}%
  \BibitemOpen
  \bibfield  {author} {\bibinfo {author} {\bibfnamefont {J.~M.}\ \bibnamefont
  {Schmidt}}, \bibinfo {author} {\bibfnamefont {G.}~\bibnamefont {Soff}},\ and\
  \bibinfo {author} {\bibfnamefont {P.~J.}\ \bibnamefont {Mohr}},\ }\href
  {https://doi.org/10.1103/PhysRevA.40.2176} {\bibfield  {journal} {\bibinfo
  {journal} {Phys. Rev. A}\ }\textbf {\bibinfo {volume} {40}},\ \bibinfo
  {pages} {2176} (\bibinfo {year} {1989})}\BibitemShut {NoStop}%
\bibitem [{\citenamefont {Lee}\ and\ \citenamefont
  {Milstein}(1994)}]{LEE_Milstein199472}%
  \BibitemOpen
  \bibfield  {author} {\bibinfo {author} {\bibfnamefont {R.}~\bibnamefont
  {Lee}}\ and\ \bibinfo {author} {\bibfnamefont {A.}~\bibnamefont {Milstein}},\
  }\href {https://doi.org/10.1016/0375-9601(94)90820-6} {\bibfield  {journal}
  {\bibinfo  {journal} {Physics Letters A}\ }\textbf {\bibinfo {volume}
  {189}},\ \bibinfo {pages} {72} (\bibinfo {year} {1994})}\BibitemShut
  {NoStop}%
\bibitem [{\citenamefont {Sapirstein}\ and\ \citenamefont
  {Cheng}(2003)}]{SapirsteinCheng2003PhysRevA.68.042111}%
  \BibitemOpen
  \bibfield  {author} {\bibinfo {author} {\bibfnamefont {J.}~\bibnamefont
  {Sapirstein}}\ and\ \bibinfo {author} {\bibfnamefont {K.~T.}\ \bibnamefont
  {Cheng}},\ }\href {https://doi.org/10.1103/PhysRevA.68.042111} {\bibfield
  {journal} {\bibinfo  {journal} {Phys. Rev. A}\ }\textbf {\bibinfo {volume}
  {68}},\ \bibinfo {pages} {042111} (\bibinfo {year} {2003})}\BibitemShut
  {NoStop}%
\bibitem [{\citenamefont {Soff}\ and\ \citenamefont
  {Mohr}(1989)}]{soff1989influence}%
  \BibitemOpen
  \bibfield  {author} {\bibinfo {author} {\bibfnamefont {G.}~\bibnamefont
  {Soff}}\ and\ \bibinfo {author} {\bibfnamefont {P.~J.}\ \bibnamefont
  {Mohr}},\ }\href {https://doi.org/10.1103/PhysRevA.40.2174} {\bibfield
  {journal} {\bibinfo  {journal} {Phys. Rev. A}\ }\textbf {\bibinfo {volume}
  {40}},\ \bibinfo {pages} {2174} (\bibinfo {year} {1989})}\BibitemShut
  {NoStop}%
\bibitem [{\citenamefont {Mohr}\ \emph {et~al.}(1998)\citenamefont {Mohr},
  \citenamefont {Plunien},\ and\ \citenamefont {Soff}}]{mohr1998qed}%
  \BibitemOpen
  \bibfield  {author} {\bibinfo {author} {\bibfnamefont {P.~J.}\ \bibnamefont
  {Mohr}}, \bibinfo {author} {\bibfnamefont {G.}~\bibnamefont {Plunien}},\ and\
  \bibinfo {author} {\bibfnamefont {G.}~\bibnamefont {Soff}},\ }\href
  {https://doi.org/10.1016/S0370-1573(97)00046-X} {\bibfield  {journal}
  {\bibinfo  {journal} {Physics Reports}\ }\textbf {\bibinfo {volume} {293}},\
  \bibinfo {pages} {227} (\bibinfo {year} {1998})}\BibitemShut {NoStop}%
\bibitem [{\citenamefont {Persson}\ \emph {et~al.}(1993)\citenamefont
  {Persson}, \citenamefont {Lindgren}, \citenamefont {Salomonson},\ and\
  \citenamefont
  {Sunnergren}}]{PersonLindgrenSalomonsonSunnergren1993PhysRevA.48.2772}%
  \BibitemOpen
  \bibfield  {author} {\bibinfo {author} {\bibfnamefont {H.}~\bibnamefont
  {Persson}}, \bibinfo {author} {\bibfnamefont {I.}~\bibnamefont {Lindgren}},
  \bibinfo {author} {\bibfnamefont {S.}~\bibnamefont {Salomonson}},\ and\
  \bibinfo {author} {\bibfnamefont {P.}~\bibnamefont {Sunnergren}},\ }\href
  {https://doi.org/10.1103/PhysRevA.48.2772} {\bibfield  {journal} {\bibinfo
  {journal} {Phys. Rev. A}\ }\textbf {\bibinfo {volume} {48}},\ \bibinfo
  {pages} {2772} (\bibinfo {year} {1993})}\BibitemShut {NoStop}%
\bibitem [{\citenamefont {Grant}\ and\ \citenamefont
  {Quiney}(2022)}]{GrantQuiney2022atoms10040108}%
  \BibitemOpen
  \bibfield  {author} {\bibinfo {author} {\bibfnamefont {I.}~\bibnamefont
  {Grant}}\ and\ \bibinfo {author} {\bibfnamefont {H.}~\bibnamefont {Quiney}},\
  }\bibfield  {journal} {\bibinfo  {journal} {Atoms}\ }\textbf {\bibinfo
  {volume} {10}},\ \href {https://doi.org/10.3390/atoms10040108}
  {10.3390/atoms10040108} (\bibinfo {year} {2022})\BibitemShut {NoStop}%
\bibitem [{\citenamefont {Sunnergren}(1998)}]{sunnergren1998Thesis}%
  \BibitemOpen
  \bibfield  {author} {\bibinfo {author} {\bibfnamefont {P.}~\bibnamefont
  {Sunnergren}},\ }\emph {\bibinfo {title} {{Complete One-Loop QED Calculations
  for Few-Electrons Ions.-Applications to Electron-Electron Interaction, the
  Zeeman Effect and Hyperfine Structure}}},\ \href
  {http://hdl.handle.net/2077/14396} {Ph.D. thesis},\ \bibinfo  {school}
  {G\"{o}teborg University} (\bibinfo {year} {1998})\BibitemShut {NoStop}%
\bibitem [{\citenamefont {Grant}(2006)}]{Grant2006_chapter}%
  \BibitemOpen
  \bibfield  {author} {\bibinfo {author} {\bibfnamefont {I.}~\bibnamefont
  {Grant}},\ }\bibinfo {title} {{Relativistic Atomic Structure}},\ in\ \href
  {https://doi.org/10.1007/978-0-387-26308-3_22} {\emph {\bibinfo {booktitle}
  {{Springer Handbook of Atomic, Molecular, and Optical Physics}}}},\ \bibinfo
  {editor} {edited by\ \bibinfo {editor} {\bibfnamefont {G.}~\bibnamefont
  {Drake}}}\ (\bibinfo  {publisher} {Springer New York},\ \bibinfo {address}
  {New York, NY},\ \bibinfo {year} {2006})\ pp.\ \bibinfo {pages}
  {325--357}\BibitemShut {NoStop}%
\bibitem [{\citenamefont {Stanton}\ and\ \citenamefont
  {Havriliak}(1984)}]{stanton1984kinetic}%
  \BibitemOpen
  \bibfield  {author} {\bibinfo {author} {\bibfnamefont {R.~E.}\ \bibnamefont
  {Stanton}}\ and\ \bibinfo {author} {\bibfnamefont {S.}~\bibnamefont
  {Havriliak}},\ }\href {https://doi.org/10.1063/1.447865} {\bibfield
  {journal} {\bibinfo  {journal} {The Journal of chemical physics}\ }\textbf
  {\bibinfo {volume} {81}},\ \bibinfo {pages} {1910} (\bibinfo {year}
  {1984})}\BibitemShut {NoStop}%
\bibitem [{\citenamefont {Furry}(1937)}]{Furry1937theorem}%
  \BibitemOpen
  \bibfield  {author} {\bibinfo {author} {\bibfnamefont {W.~H.}\ \bibnamefont
  {Furry}},\ }\href {https://doi.org/10.1103/PhysRev.51.125} {\bibfield
  {journal} {\bibinfo  {journal} {Phys. Rev.}\ }\textbf {\bibinfo {volume}
  {51}},\ \bibinfo {pages} {125} (\bibinfo {year} {1937})}\BibitemShut
  {NoStop}%
\bibitem [{\citenamefont {Schwinger}(1951)}]{SchwingerPhysRev.82.664_1951}%
  \BibitemOpen
  \bibfield  {author} {\bibinfo {author} {\bibfnamefont {J.}~\bibnamefont
  {Schwinger}},\ }\href {https://doi.org/10.1103/PhysRev.82.664} {\bibfield
  {journal} {\bibinfo  {journal} {Phys. Rev.}\ }\textbf {\bibinfo {volume}
  {82}},\ \bibinfo {pages} {664} (\bibinfo {year} {1951})}\BibitemShut
  {NoStop}%
\bibitem [{\citenamefont {Feynman}(1949)}]{Feynamn_1949_Positron}%
  \BibitemOpen
  \bibfield  {author} {\bibinfo {author} {\bibfnamefont {R.~P.}\ \bibnamefont
  {Feynman}},\ }\href {https://doi.org/10.1103/PhysRev.76.749} {\bibfield
  {journal} {\bibinfo  {journal} {Phys. Rev.}\ }\textbf {\bibinfo {volume}
  {76}},\ \bibinfo {pages} {749} (\bibinfo {year} {1949})}\BibitemShut
  {NoStop}%
\bibitem [{\citenamefont {Itzykson}\ and\ \citenamefont
  {Zuber}(1980)}]{Itzykson_Zuber_QFT_1980}%
  \BibitemOpen
  \bibfield  {author} {\bibinfo {author} {\bibfnamefont {C.}~\bibnamefont
  {Itzykson}}\ and\ \bibinfo {author} {\bibfnamefont {J.}~\bibnamefont
  {Zuber}},\ }\href {https://store.doverpublications.com/0486445682.html}
  {\emph {\bibinfo {title} {{Quantum Field Theory}}}},\ Dover Books on Physics\
  (\bibinfo  {publisher} {Dover Publications},\ \bibinfo {year}
  {1980})\BibitemShut {NoStop}%
\bibitem [{\citenamefont {Schweber}(2011)}]{schweber2011introduction}%
  \BibitemOpen
  \bibfield  {author} {\bibinfo {author} {\bibfnamefont {S.~S.}\ \bibnamefont
  {Schweber}},\ }\href@noop {} {\emph {\bibinfo {title} {{An Introduction to
  Relativistic Quantum Field Theory}}}}\ (\bibinfo  {publisher} {Dover
  Publications},\ \bibinfo {year} {2011})\BibitemShut {NoStop}%
\bibitem [{\citenamefont {Greiner}\ and\ \citenamefont
  {Reinhardt}(2009)}]{greiner_reinhard2009QED}%
  \BibitemOpen
  \bibfield  {author} {\bibinfo {author} {\bibfnamefont {W.}~\bibnamefont
  {Greiner}}\ and\ \bibinfo {author} {\bibfnamefont {J.}~\bibnamefont
  {Reinhardt}},\ }\href {https://doi.org/10.1007/978-3-540-87561-1} {\emph
  {\bibinfo {title} {{Quantum Electrodynamics}}}},\ \bibinfo {edition} {4th}\
  ed.\ (\bibinfo  {publisher} {Springer-Verlag Berlin Heidelberg},\ \bibinfo
  {year} {2009})\BibitemShut {NoStop}%
\bibitem [{\citenamefont {Greiner}\ \emph {et~al.}(1985)\citenamefont
  {Greiner}, \citenamefont {M{\"u}ller},\ and\ \citenamefont
  {Rafelski}}]{Greiner1985QEDstrong}%
  \BibitemOpen
  \bibfield  {author} {\bibinfo {author} {\bibfnamefont {W.}~\bibnamefont
  {Greiner}}, \bibinfo {author} {\bibfnamefont {B.}~\bibnamefont
  {M{\"u}ller}},\ and\ \bibinfo {author} {\bibfnamefont {J.}~\bibnamefont
  {Rafelski}},\ }\href {https://doi.org/10.1007/978-3-642-82272-8} {\emph
  {\bibinfo {title} {{Quantum Electrodynamics of Strong Fields}}}}\ (\bibinfo
  {publisher} {Springer Berlin Heidelberg},\ \bibinfo {year}
  {1985})\BibitemShut {NoStop}%
\bibitem [{\citenamefont {Peierls}(1934)}]{peierls1934vacuum}%
  \BibitemOpen
  \bibfield  {author} {\bibinfo {author} {\bibfnamefont {R.}~\bibnamefont
  {Peierls}},\ }\href {https://doi.org/10.1098/rspa.1934.0164} {\bibfield
  {journal} {\bibinfo  {journal} {Proceedings of the Royal Society of London}\
  }\textbf {\bibinfo {volume} {146}},\ \bibinfo {pages} {420} (\bibinfo {year}
  {1934})}\BibitemShut {NoStop}%
\bibitem [{\citenamefont {Indelicato}\ \emph {et~al.}(2014)\citenamefont
  {Indelicato}, \citenamefont {Mohr},\ and\ \citenamefont
  {Sapirstein}}]{Indelicato_Mohr_SapirsteinCoordinateSpaceVP2014}%
  \BibitemOpen
  \bibfield  {author} {\bibinfo {author} {\bibfnamefont {P.}~\bibnamefont
  {Indelicato}}, \bibinfo {author} {\bibfnamefont {P.~J.}\ \bibnamefont
  {Mohr}},\ and\ \bibinfo {author} {\bibfnamefont {J.}~\bibnamefont
  {Sapirstein}},\ }\href {https://doi.org/10.1103/PhysRevA.89.042121}
  {\bibfield  {journal} {\bibinfo  {journal} {Phys. Rev. A}\ }\textbf {\bibinfo
  {volume} {89}},\ \bibinfo {pages} {042121} (\bibinfo {year}
  {2014})}\BibitemShut {NoStop}%
\bibitem [{\citenamefont {Salman}(2022)}]{salman2022quantum}%
  \BibitemOpen
  \bibfield  {author} {\bibinfo {author} {\bibfnamefont {M.}~\bibnamefont
  {Salman}},\ }\emph {\bibinfo {title} {{Quantum electrodynamic corrections in
  quantum chemistry}}},\ \href
  {https://ut3-toulouseinp.hal.science/tel-03715663/} {Ph.D. thesis},\ \bibinfo
   {school} {Universit{\'e} Paul Sabatier-Toulouse III} (\bibinfo {year}
  {2022})\BibitemShut {NoStop}%
\bibitem [{\citenamefont {Schwabl}(2008)}]{schwabl2007quantum}%
  \BibitemOpen
  \bibfield  {author} {\bibinfo {author} {\bibfnamefont {F.}~\bibnamefont
  {Schwabl}},\ }\href {https://doi.org/10.1007/978-3-540-85062-5} {\emph
  {\bibinfo {title} {Advanced Quantum Mechanics}}},\ \bibinfo {edition} {4th}\
  ed.\ (\bibinfo  {publisher} {Springer Berlin, Heidelberg},\ \bibinfo {year}
  {2008})\BibitemShut {NoStop}%
\bibitem [{\citenamefont {Dirac}(1934)}]{dirac_1934}%
  \BibitemOpen
  \bibfield  {author} {\bibinfo {author} {\bibfnamefont {P.~A.~M.}\
  \bibnamefont {Dirac}},\ }\href {https://doi.org/10.1017/S030500410001656X}
  {\bibfield  {journal} {\bibinfo  {journal} {Mathematical Proceedings of the
  Cambridge Philosophical Society}\ }\textbf {\bibinfo {volume} {30}},\
  \bibinfo {pages} {150} (\bibinfo {year} {1934})}\BibitemShut {NoStop}%
\bibitem [{\citenamefont {Schwinger}(1948)}]{SchwingerPhysRev.74.1439_1948}%
  \BibitemOpen
  \bibfield  {author} {\bibinfo {author} {\bibfnamefont {J.}~\bibnamefont
  {Schwinger}},\ }\href {https://doi.org/10.1103/PhysRev.74.1439} {\bibfield
  {journal} {\bibinfo  {journal} {Phys. Rev.}\ }\textbf {\bibinfo {volume}
  {74}},\ \bibinfo {pages} {1439} (\bibinfo {year} {1948})}\BibitemShut
  {NoStop}%
\bibitem [{\citenamefont {Schwinger}(1949)}]{SchwingerPhysRev.75.651_1949}%
  \BibitemOpen
  \bibfield  {author} {\bibinfo {author} {\bibfnamefont {J.}~\bibnamefont
  {Schwinger}},\ }\href {https://doi.org/10.1103/PhysRev.75.651} {\bibfield
  {journal} {\bibinfo  {journal} {Phys. Rev.}\ }\textbf {\bibinfo {volume}
  {75}},\ \bibinfo {pages} {651} (\bibinfo {year} {1949})}\BibitemShut
  {NoStop}%
\bibitem [{\citenamefont {Johnson}(2007)}]{johnson2007atomic}%
  \BibitemOpen
  \bibfield  {author} {\bibinfo {author} {\bibfnamefont {W.~R.}\ \bibnamefont
  {Johnson}},\ }\href {https://doi.org/10.1007/978-3-540-68013-0} {\emph
  {\bibinfo {title} {{Atomic structure theory}}}}\ (\bibinfo  {publisher}
  {Springer-Verlag Berlin Heidelberg},\ \bibinfo {year} {2007})\BibitemShut
  {NoStop}%
\bibitem [{\citenamefont {Szmytkowski}(2005)}]{Szmytkowski_2005}%
  \BibitemOpen
  \bibfield  {author} {\bibinfo {author} {\bibfnamefont {R.}~\bibnamefont
  {Szmytkowski}},\ }\href {https://doi.org/10.1088/0305-4470/38/41/011}
  {\bibfield  {journal} {\bibinfo  {journal} {Journal of Physics A:
  Mathematical and General}\ }\textbf {\bibinfo {volume} {38}},\ \bibinfo
  {pages} {8993} (\bibinfo {year} {2005})}\BibitemShut {NoStop}%
\bibitem [{\citenamefont {Salman}\ and\ \citenamefont
  {Saue}(2020)}]{Csymm_salman_saue}%
  \BibitemOpen
  \bibfield  {author} {\bibinfo {author} {\bibfnamefont {M.}~\bibnamefont
  {Salman}}\ and\ \bibinfo {author} {\bibfnamefont {T.}~\bibnamefont {Saue}},\
  }\bibfield  {journal} {\bibinfo  {journal} {Symmetry}\ }\textbf {\bibinfo
  {volume} {12}},\ \href {https://doi.org/10.3390/sym12071121}
  {10.3390/sym12071121} (\bibinfo {year} {2020})\BibitemShut {NoStop}%
\bibitem [{\citenamefont {Kutzelnigg}(1984)}]{kutzelnigg1984basis}%
  \BibitemOpen
  \bibfield  {author} {\bibinfo {author} {\bibfnamefont {W.}~\bibnamefont
  {Kutzelnigg}},\ }\href {https://doi.org/10.1002/QUA.560250112} {\bibfield
  {journal} {\bibinfo  {journal} {International Journal of Quantum Chemistry}\
  }\textbf {\bibinfo {volume} {25}},\ \bibinfo {pages} {107} (\bibinfo {year}
  {1984})}\BibitemShut {NoStop}%
\bibitem [{\citenamefont {Tupitsyn}\ and\ \citenamefont
  {Shabaev}(2008)}]{tupitsyn_Shabaev_2008spurious}%
  \BibitemOpen
  \bibfield  {author} {\bibinfo {author} {\bibfnamefont {I.}~\bibnamefont
  {Tupitsyn}}\ and\ \bibinfo {author} {\bibfnamefont {V.}~\bibnamefont
  {Shabaev}},\ }\href {https://doi.org/10.1134/S0030400X08080043} {\bibfield
  {journal} {\bibinfo  {journal} {Optics and Spectroscopy}\ }\textbf {\bibinfo
  {volume} {105}},\ \bibinfo {pages} {183} (\bibinfo {year}
  {2008})}\BibitemShut {NoStop}%
\bibitem [{\citenamefont {Goldman}(1985)}]{Goldman1985}%
  \BibitemOpen
  \bibfield  {author} {\bibinfo {author} {\bibfnamefont {S.~P.}\ \bibnamefont
  {Goldman}},\ }\href {https://doi.org/10.1103/PhysRevA.31.3541} {\bibfield
  {journal} {\bibinfo  {journal} {Phys. Rev. A}\ }\textbf {\bibinfo {volume}
  {31}},\ \bibinfo {pages} {3541} (\bibinfo {year} {1985})}\BibitemShut
  {NoStop}%
\bibitem [{\citenamefont {Lewin}\ and\ \citenamefont
  {S{\'e}r{\'e}}(2009)}]{Lewin_Sere_2009_spectral-pollution}%
  \BibitemOpen
  \bibfield  {author} {\bibinfo {author} {\bibfnamefont {M.}~\bibnamefont
  {Lewin}}\ and\ \bibinfo {author} {\bibfnamefont {{\'E}.}~\bibnamefont
  {S{\'e}r{\'e}}},\ }\href {https://doi.org/10.1112/plms/pdp046} {\bibfield
  {journal} {\bibinfo  {journal} {Proceedings of the London Mathematical
  Society}\ }\textbf {\bibinfo {volume} {100}},\ \bibinfo {pages} {864}
  (\bibinfo {year} {2009})}\BibitemShut {NoStop}%
\bibitem [{\citenamefont {Lewin}\ and\ \citenamefont
  {S{\'e}r{\'e}}(2014)}]{lewin2014spurious}%
  \BibitemOpen
  \bibfield  {author} {\bibinfo {author} {\bibfnamefont {M.}~\bibnamefont
  {Lewin}}\ and\ \bibinfo {author} {\bibfnamefont {{\'E}.}~\bibnamefont
  {S{\'e}r{\'e}}},\ }\bibinfo {title} {{Spurious Modes in Dirac Calculations
  and How to Avoid Them}},\ in\ \href
  {https://doi.org/10.1007/978-3-319-06379-9_2} {\emph {\bibinfo {booktitle}
  {{Many-Electron Approaches in Physics, Chemistry and Mathematics: A
  Multidisciplinary View}}}},\ \bibinfo {editor} {edited by\ \bibinfo {editor}
  {\bibfnamefont {V.}~\bibnamefont {Bach}}\ and\ \bibinfo {editor}
  {\bibfnamefont {L.}~\bibnamefont {Delle~Site}}}\ (\bibinfo  {publisher}
  {Springer International Publishing},\ \bibinfo {address} {Cham},\ \bibinfo
  {year} {2014})\ pp.\ \bibinfo {pages} {31--52}\BibitemShut {NoStop}%
\bibitem [{\citenamefont {Schwarz}\ and\ \citenamefont
  {Wallmeier}(1982)}]{Schwarz_and_Wallmeir_1982}%
  \BibitemOpen
  \bibfield  {author} {\bibinfo {author} {\bibfnamefont {W.}~\bibnamefont
  {Schwarz}}\ and\ \bibinfo {author} {\bibfnamefont {H.}~\bibnamefont
  {Wallmeier}},\ }\href {https://doi.org/10.1080/00268978200101771} {\bibfield
  {journal} {\bibinfo  {journal} {Molecular Physics}\ }\textbf {\bibinfo
  {volume} {46}},\ \bibinfo {pages} {1045} (\bibinfo {year}
  {1982})}\BibitemShut {NoStop}%
\bibitem [{\citenamefont {Grant}(1982)}]{Grant_1982}%
  \BibitemOpen
  \bibfield  {author} {\bibinfo {author} {\bibfnamefont {I.~P.}\ \bibnamefont
  {Grant}},\ }\href {https://doi.org/10.1103/PhysRevA.25.1230} {\bibfield
  {journal} {\bibinfo  {journal} {Phys. Rev. A}\ }\textbf {\bibinfo {volume}
  {25}},\ \bibinfo {pages} {1230} (\bibinfo {year} {1982})}\BibitemShut
  {NoStop}%
\bibitem [{\citenamefont {Dyall}\ \emph {et~al.}(1984)\citenamefont {Dyall},
  \citenamefont {Grant},\ and\ \citenamefont
  {Wilson}}]{Dyall_1984_matrix_operators}%
  \BibitemOpen
  \bibfield  {author} {\bibinfo {author} {\bibfnamefont {K.~G.}\ \bibnamefont
  {Dyall}}, \bibinfo {author} {\bibfnamefont {I.~P.}\ \bibnamefont {Grant}},\
  and\ \bibinfo {author} {\bibfnamefont {S.}~\bibnamefont {Wilson}},\ }\href
  {https://doi.org/10.1088/0022-3700/17/4/006} {\bibfield  {journal} {\bibinfo
  {journal} {Journal of Physics B: Atomic and Molecular Physics}\ }\textbf
  {\bibinfo {volume} {17}},\ \bibinfo {pages} {493} (\bibinfo {year}
  {1984})}\BibitemShut {NoStop}%
\bibitem [{\citenamefont {Kutzelnigg}(2007)}]{kutzelnigg2007completeness}%
  \BibitemOpen
  \bibfield  {author} {\bibinfo {author} {\bibfnamefont {W.}~\bibnamefont
  {Kutzelnigg}},\ }\href {https://doi.org/10.1063/1.2744018} {\bibfield
  {journal} {\bibinfo  {journal} {The Journal of chemical physics}\ }\textbf
  {\bibinfo {volume} {126}},\ \bibinfo {pages} {201103} (\bibinfo {year}
  {2007})}\BibitemShut {NoStop}%
\bibitem [{\citenamefont {Sun}\ \emph {et~al.}(2011)\citenamefont {Sun},
  \citenamefont {Liu},\ and\ \citenamefont {Kutzelnigg}}]{sun2011comparison}%
  \BibitemOpen
  \bibfield  {author} {\bibinfo {author} {\bibfnamefont {Q.}~\bibnamefont
  {Sun}}, \bibinfo {author} {\bibfnamefont {W.}~\bibnamefont {Liu}},\ and\
  \bibinfo {author} {\bibfnamefont {W.}~\bibnamefont {Kutzelnigg}},\ }\href
  {https://doi.org/10.1007/s00214-010-0876-6} {\bibfield  {journal} {\bibinfo
  {journal} {Theoretical Chemistry Accounts}\ }\textbf {\bibinfo {volume}
  {129}},\ \bibinfo {pages} {423} (\bibinfo {year} {2011})}\BibitemShut
  {NoStop}%
\bibitem [{\citenamefont {Shabaev}\ \emph {et~al.}(2004)\citenamefont
  {Shabaev}, \citenamefont {Tupitsyn}, \citenamefont {Yerokhin}, \citenamefont
  {Plunien},\ and\ \citenamefont {Soff}}]{Shabaev_2004_DKB}%
  \BibitemOpen
  \bibfield  {author} {\bibinfo {author} {\bibfnamefont {V.~M.}\ \bibnamefont
  {Shabaev}}, \bibinfo {author} {\bibfnamefont {I.~I.}\ \bibnamefont
  {Tupitsyn}}, \bibinfo {author} {\bibfnamefont {V.~A.}\ \bibnamefont
  {Yerokhin}}, \bibinfo {author} {\bibfnamefont {G.}~\bibnamefont {Plunien}},\
  and\ \bibinfo {author} {\bibfnamefont {G.}~\bibnamefont {Soff}},\ }\href
  {https://doi.org/10.1103/PhysRevLett.93.130405} {\bibfield  {journal}
  {\bibinfo  {journal} {Phys. Rev. Lett.}\ }\textbf {\bibinfo {volume} {93}},\
  \bibinfo {pages} {130405} (\bibinfo {year} {2004})}\BibitemShut {NoStop}%
\bibitem [{\citenamefont {Dyall}\ and\ \citenamefont
  {F{\ae}gri}(1996)}]{DyallandFaegri1996}%
  \BibitemOpen
  \bibfield  {author} {\bibinfo {author} {\bibfnamefont {K.~G.}\ \bibnamefont
  {Dyall}}\ and\ \bibinfo {author} {\bibfnamefont {K.}~\bibnamefont
  {F{\ae}gri}},\ }\href {https://doi.org/10.1007/BF00190154} {\bibfield
  {journal} {\bibinfo  {journal} {Theoretica chimica acta}\ }\textbf {\bibinfo
  {volume} {94}},\ \bibinfo {pages} {39} (\bibinfo {year} {1996})}\BibitemShut
  {NoStop}%
\bibitem [{\citenamefont {Abramowitz}\ and\ \citenamefont
  {Stegun}(1972)}]{Abramowitz+stegun}%
  \BibitemOpen
  \bibinfo {editor} {\bibfnamefont {M.}~\bibnamefont {Abramowitz}}\ and\
  \bibinfo {editor} {\bibfnamefont {I.~A.}\ \bibnamefont {Stegun}},\ eds.,\
  \href@noop {} {\emph {\bibinfo {title} {{Handbook of Mathematical Functions
  with Formulas, Graphs, and Mathematical Tables}}}},\ \bibinfo {edition}
  {10th}\ ed.\ (\bibinfo  {publisher} {National Bureau of Standards (became:
  National Institute of Standards and Technology)},\ \bibinfo {year}
  {1972})\BibitemShut {NoStop}%
\bibitem [{\citenamefont {Fullerton}\ and\ \citenamefont
  {Rinker}(1976)}]{Fullerton_Rinker_1976}%
  \BibitemOpen
  \bibfield  {author} {\bibinfo {author} {\bibfnamefont {L.~W.}\ \bibnamefont
  {Fullerton}}\ and\ \bibinfo {author} {\bibfnamefont {G.~A.}\ \bibnamefont
  {Rinker}},\ }\href {https://doi.org/10.1103/PhysRevA.13.1283} {\bibfield
  {journal} {\bibinfo  {journal} {Phys. Rev. A}\ }\textbf {\bibinfo {volume}
  {13}},\ \bibinfo {pages} {1283} (\bibinfo {year} {1976})}\BibitemShut
  {NoStop}%
\bibitem [{\citenamefont {Tiesinga}\ \emph {et~al.}(2021)\citenamefont
  {Tiesinga}, \citenamefont {Mohr}, \citenamefont {Newell},\ and\ \citenamefont
  {Taylor}}]{2021codata2018}%
  \BibitemOpen
  \bibfield  {author} {\bibinfo {author} {\bibfnamefont {E.}~\bibnamefont
  {Tiesinga}}, \bibinfo {author} {\bibfnamefont {P.~J.}\ \bibnamefont {Mohr}},
  \bibinfo {author} {\bibfnamefont {D.~B.}\ \bibnamefont {Newell}},\ and\
  \bibinfo {author} {\bibfnamefont {B.~N.}\ \bibnamefont {Taylor}},\ }\href
  {https://doi.org/10.1063/5.0064853} {\bibfield  {journal} {\bibinfo
  {journal} {Journal of Physical and Chemical Reference Data}\ }\textbf
  {\bibinfo {volume} {50}},\ \bibinfo {pages} {033105} (\bibinfo {year}
  {2021})}\BibitemShut {NoStop}%
\bibitem [{\citenamefont {{Wolfram Research{,} Inc.}}()}]{Mathematica}%
  \BibitemOpen
  \bibfield  {author} {\bibinfo {author} {\bibnamefont {{Wolfram Research{,}
  Inc.}}},\ }\href {https://www.wolfram.com/mathematica/} {\bibinfo {title}
  {{Mathematica}}},\ \bibinfo {note} {2020}\BibitemShut {NoStop}%
\bibitem [{\citenamefont {Visscher}\ and\ \citenamefont
  {Dyall}(1997)}]{VISSCHER1997207}%
  \BibitemOpen
  \bibfield  {author} {\bibinfo {author} {\bibfnamefont {L.}~\bibnamefont
  {Visscher}}\ and\ \bibinfo {author} {\bibfnamefont {K.}~\bibnamefont
  {Dyall}},\ }\href {https://doi.org/10.1006/adnd.1997.0751} {\bibfield
  {journal} {\bibinfo  {journal} {Atomic Data and Nuclear Data Tables}\
  }\textbf {\bibinfo {volume} {67}},\ \bibinfo {pages} {207} (\bibinfo {year}
  {1997})}\BibitemShut {NoStop}%
\bibitem [{\citenamefont {Andrae}(2000)}]{Andrae2000}%
  \BibitemOpen
  \bibfield  {author} {\bibinfo {author} {\bibfnamefont {D.}~\bibnamefont
  {Andrae}},\ }\href {https://doi.org//10.1016/S0370-1573(00)00007-7}
  {\bibfield  {journal} {\bibinfo  {journal} {Phys. Rep.}\ }\textbf {\bibinfo
  {volume} {336}},\ \bibinfo {pages} {413} (\bibinfo {year}
  {2000})}\BibitemShut {NoStop}%
\bibitem [{\citenamefont {Norman}\ \emph {et~al.}(2018)\citenamefont {Norman},
  \citenamefont {Ruud},\ and\ \citenamefont {Saue}}]{nrsbook}%
  \BibitemOpen
  \bibfield  {author} {\bibinfo {author} {\bibfnamefont {P.}~\bibnamefont
  {Norman}}, \bibinfo {author} {\bibfnamefont {K.}~\bibnamefont {Ruud}},\ and\
  \bibinfo {author} {\bibfnamefont {T.}~\bibnamefont {Saue}},\ }\href
  {https://doi.org/10.1002/9781118794821} {\emph {\bibinfo {title} {{Principles
  and Practices of Molecular Properties: Theory, Modeling and Simulations}}}}\
  (\bibinfo  {publisher} {Wiley},\ \bibinfo {address} {Hoboken, NJ},\ \bibinfo
  {year} {2018})\BibitemShut {NoStop}%
\bibitem [{\citenamefont {Froese~Fischer}\ \emph {et~al.}(1997)\citenamefont
  {Froese~Fischer}, \citenamefont {Barge},\ and\ \citenamefont
  {J\"{o}nsson}}]{froese1997computational}%
  \BibitemOpen
  \bibfield  {author} {\bibinfo {author} {\bibfnamefont {C.}~\bibnamefont
  {Froese~Fischer}}, \bibinfo {author} {\bibfnamefont {T.}~\bibnamefont
  {Barge}},\ and\ \bibinfo {author} {\bibfnamefont {P.}~\bibnamefont
  {J\"{o}nsson}},\ }\href {https://doi.org/10.1201/9781315139982} {\emph
  {\bibinfo {title} {{Computational atomic structure: an MCHF approach}}}}\
  (\bibinfo  {publisher} {CRC press, Taylor \& Francis},\ \bibinfo {year}
  {1997})\BibitemShut {NoStop}%
\bibitem [{\citenamefont {Almoukhalalati}\ \emph {et~al.}(2016)\citenamefont
  {Almoukhalalati}, \citenamefont {Knecht}, \citenamefont {Jensen},
  \citenamefont {Dyall},\ and\ \citenamefont
  {Saue}}]{almoukhalalati2016electron}%
  \BibitemOpen
  \bibfield  {author} {\bibinfo {author} {\bibfnamefont {A.}~\bibnamefont
  {Almoukhalalati}}, \bibinfo {author} {\bibfnamefont {S.}~\bibnamefont
  {Knecht}}, \bibinfo {author} {\bibfnamefont {H.~J.~{\relax Aa}.}\
  \bibnamefont {Jensen}}, \bibinfo {author} {\bibfnamefont {K.~G.}\
  \bibnamefont {Dyall}},\ and\ \bibinfo {author} {\bibfnamefont
  {T.}~\bibnamefont {Saue}},\ }\href {https://doi.org/10.1063/1.4959452}
  {\bibfield  {journal} {\bibinfo  {journal} {The Journal of chemical physics}\
  }\textbf {\bibinfo {volume} {145}},\ \bibinfo {pages} {074104} (\bibinfo
  {year} {2016})}\BibitemShut {NoStop}%
\bibitem [{\citenamefont {Johnson}\ and\ \citenamefont
  {Soff}(1985)}]{JOHNSON1985405}%
  \BibitemOpen
  \bibfield  {author} {\bibinfo {author} {\bibfnamefont {W.}~\bibnamefont
  {Johnson}}\ and\ \bibinfo {author} {\bibfnamefont {G.}~\bibnamefont {Soff}},\
  }\href {https://doi.org/10.1016/0092-640X(85)90010-5} {\bibfield  {journal}
  {\bibinfo  {journal} {Atomic Data and Nuclear Data Tables}\ }\textbf
  {\bibinfo {volume} {33}},\ \bibinfo {pages} {405} (\bibinfo {year}
  {1985})}\BibitemShut {NoStop}%
\bibitem [{\citenamefont {Ishikawa}\ \emph {et~al.}(1985)\citenamefont
  {Ishikawa}, \citenamefont {Baretty},\ and\ \citenamefont
  {Binning~Jr}}]{ISHIKAWA1985130}%
  \BibitemOpen
  \bibfield  {author} {\bibinfo {author} {\bibfnamefont {Y.}~\bibnamefont
  {Ishikawa}}, \bibinfo {author} {\bibfnamefont {R.}~\bibnamefont {Baretty}},\
  and\ \bibinfo {author} {\bibfnamefont {R.}~\bibnamefont {Binning~Jr}},\
  }\href {https://doi.org/10.1016/0009-2614(85)87169-4} {\bibfield  {journal}
  {\bibinfo  {journal} {Chemical Physics Letters}\ }\textbf {\bibinfo {volume}
  {121}},\ \bibinfo {pages} {130} (\bibinfo {year} {1985})}\BibitemShut
  {NoStop}%
\bibitem [{\citenamefont {Boys}(1950)}]{boys1950electronic}%
  \BibitemOpen
  \bibfield  {author} {\bibinfo {author} {\bibfnamefont {S.~F.}\ \bibnamefont
  {Boys}},\ }\href {https://doi.org/10.1098/rspa.1950.0036} {\bibfield
  {journal} {\bibinfo  {journal} {Proceedings of the Royal Society of London.
  Series A.}\ }\textbf {\bibinfo {volume} {200}},\ \bibinfo {pages} {542}
  (\bibinfo {year} {1950})}\BibitemShut {NoStop}%
\bibitem [{\citenamefont {Szabo}\ and\ \citenamefont
  {Ostlund}(1996)}]{szabo_Ostlund_2012modern}%
  \BibitemOpen
  \bibfield  {author} {\bibinfo {author} {\bibfnamefont {A.}~\bibnamefont
  {Szabo}}\ and\ \bibinfo {author} {\bibfnamefont {N.~S.}\ \bibnamefont
  {Ostlund}},\ }\href {https://books.google.fr/books?id=KQ3DAgAAQBAJ} {\emph
  {\bibinfo {title} {{Modern quantum chemistry: introduction to advanced
  electronic structure theory}}}}\ (\bibinfo  {publisher} {Dover
  Publications},\ \bibinfo {year} {1996})\BibitemShut {NoStop}%
\bibitem [{\citenamefont {Helgaker}\ \emph {et~al.}(2000)\citenamefont
  {Helgaker}, \citenamefont {J{\o}rgensen},\ and\ \citenamefont
  {Olsen}}]{helgaker2014molecular}%
  \BibitemOpen
  \bibfield  {author} {\bibinfo {author} {\bibfnamefont {T.}~\bibnamefont
  {Helgaker}}, \bibinfo {author} {\bibfnamefont {P.}~\bibnamefont
  {J{\o}rgensen}},\ and\ \bibinfo {author} {\bibfnamefont {J.}~\bibnamefont
  {Olsen}},\ }\href {https://doi.org/10.1002/9781119019572} {\emph {\bibinfo
  {title} {{Molecular electronic-structure theory}}}}\ (\bibinfo  {publisher}
  {John Wiley \& Sons},\ \bibinfo {year} {2000})\BibitemShut {NoStop}%
\bibitem [{\citenamefont {Dyall}\ and\ \citenamefont
  {F{\ae}gri}(2007)}]{DyallFaegriRQC2007}%
  \BibitemOpen
  \bibfield  {author} {\bibinfo {author} {\bibfnamefont {K.~G.}\ \bibnamefont
  {Dyall}}\ and\ \bibinfo {author} {\bibfnamefont {K.}~\bibnamefont
  {F{\ae}gri}},\ }\href {https://doi.org/10.1093/oso/9780195140866.001.0001}
  {\emph {\bibinfo {title} {{Introduction to Relativistic Quantum
  Chemistry}}}}\ (\bibinfo  {publisher} {Oxford University Press},\ \bibinfo
  {year} {2007})\BibitemShut {NoStop}%
\end{thebibliography}%

\end{document}